\documentclass[useAMS,usenatbib]{mn2e}
\usepackage{graphicx}
\usepackage{epstopdf}
\usepackage{amsmath}    
\usepackage{bm}
\usepackage{txfonts}
\usepackage{xspace}
\usepackage{natbib}
\bibpunct{(}{)}{;}{a}{}{,}

\newcommand{\FermiLAT}{\textit{Fermi}-LAT\xspace}

\newcommand{\He}{H.E.S.S.\xspace}
\newcommand{\Ve}{VERITAS\xspace}
\newcommand{\HeII}{H.E.S.S.~II\xspace}
\newcommand{\Ma}{MAGIC\xspace}
\newcommand{\MaII}{MAGIC-II\xspace}

\title[Constraints on the SFRD from the EBL]{Probing the peak of the star formation rate density with the extragalactic background light
}

\author[M. Raue and M. Meyer]{Martin Raue$^{1}$\thanks{E-mail:martin.raue@desy.de} and Manuel Meyer$^{1}$\\
$^{1}$Institut f\"ur Experimentalphysik, Universit\"at Hamburg, Luruper Chaussee 149, 22761 Hamburg, Germany}

\begin{document}

\maketitle

\label{firstpage}

\begin{abstract}
The extragalactic background light (EBL), i.e., the diffuse meta-galactic photon field in the ultraviolet to infrared, is dominated by the emission from stars in galaxies. It is, therefore, intimately connected with the integrated star formation rate density (SFRD).
In this paper, the SFRD is constrained using recent limits on the EBL density derived from observations of distant sources of high and very-high energy gamma-rays. The stellar EBL contribution is modeled utilizing simple stellar population spectra including dust attenuation and emission.
{
For modeling the SFRD up to $z=4$ a broken power law function in $z+1$ is assumed.
}
A wide range of values for the different model parameters (SFRD($z$), metallicity, dust absorption) is investigated and their impact on the resulting EBL is studied. 
The calculated EBL densities are compared with the specific EBL density limits and constraints on the SFRD are derived.
For the fiducial model, adopting a Chabrier initial mass function (IMF)
{
and a second power law index for the SFRD of $\beta=0.3$,
}
the SFRD is constrained to $\lesssim 0.1\,\text{M}_\odot\,\text{yr}^{-1}\,\text{Mpc}^{-3}$ and $< 0.2\,\text{M}_\odot\,\text{yr}^{-1}\,\text{Mpc}^{-3}$ for a redshift of $z\sim1$ and $z\sim2$, respectively.
{
The limits for a redshift of $z\sim1$ are in tension with SFRD measurements derived from instantaneous star formation tracers. 
}
While the tension for the conservative fiducial model in this study is not yet overly strong, the tension increases when applying plausible changes to the model parameters, e.g., using a Salpeter instead of a Chabrier IMF or a adopting a sub-solar metallicity.
\end{abstract}



\section{Introduction}

The star formation rate density (SFRD) describes the evolution of stellar formation over the history of the universe and is closely connected to structure formation and reioniziation \citep[for a recent review see][]{robertson:2010a}. The SFRD can be obtained by combining the star formation rate for individual galaxies as derived, e.g., from the total infrared (IR), radio luminosity, or the Lyman-$\alpha$ emission \citep{kennicutt:1998a}, with the space density of the sources \citep[see][for recent data compilations and references]{hopkins:2004a, hopkins:2006a}.

The extragalactic background light \citep[EBL; see][for an excellent review]{hauser:2001a}, which is the diffuse meta-galactic photon field in the ultraviolet (UV) to the infrared (IR) wavelength regime, is dominated by stellar emission in the optical (O) to near-infrared (NIR) and by stellar emission reprocessed by dust in the mid to far-infrared (MIR/FIR). It, therefore, provides a probe of the integrated SFRD.
Direct measurements of the EBL are challenging due to strong foreground emission in our planetary system (zodiacal light) and galaxy \citep{hauser:1998a}. Lower limits on its density are derived from integrated source counts \citep[e.g.][]{madau:2000a, fazio:2004a, dole:2006a}.

The most constraining upper limits on the EBL density are obtained from spectroscopic observations of distant sources of very-high energy gamma-rays (VHE; $E>100\,\text{GeV}$\footnote{$1\,\text{MeV} = 10^6\,\text{eV},1\,\text{GeV} = 10^9\,\text{eV}, 1\,\text{TeV} = 10^{12}\,\text{eV}$}): VHE gamma-rays produce electron-positron pairs with the low energy photons of the EBL ($\gamma_\text{VHE}\,\gamma_\text{EBL}\rightarrow{}e^-\,e^+$), resulting in an energy dependent attenuation signature in the measured VHE gamma-ray spectra \citep{nikishov:1962a,jelley:1966a,gould:1967a}. Using assumptions about the intrinsic spectrum emitted at the source the EBL density can be constrained \citep{stecker:1992a}. The discovery of distant ($z\sim0.2-0.5$) sources of VHE gamma-rays with current generation ground-based VHE instruments like \He, \Ma, and \Ve lead to strong constraints on the EBL density, in particular in the O to NIR \citep{aharonian:2006:hess:ebl:nature,albert:2008:magic:3c279:science}. These instruments also increased the number of known extragalactic VHE sources from about five in the year 2000 to more then 50 in the year 2011\footnote{\texttt{http://tevcat.uchicago.edu/}}. By combining the observations of several VHE sources, the systematic uncertainties in deriving upper limits on the EBL density can be significantly decreased and constraints over a wide wavelength range of the EBL can be derived \citep{dwek:2005a,mazin:2007a}. In particular, in a recent study by \citet[ME12 in the following]{meyer:2012a} data from the \FermiLAT instrument, covering the MeV to GeV energy regime, were combined with a large sample of extragalactic VHE sources to derive the strongest and most robust upper limits on the EBL density today.

Limits and measurements of the EBL can be used to investigate the SFRD in the early universe \citep{santos:2002a, fernandez:2006a, raue:2009a, maurer:2012a, gilmore:2012a}. 
{
In principle such studies provide a way of investigating the aggregated star formation of galaxies too faint to detect individually with current technology.
}
Unfortunately, the current limits are not strongly constraining, due to the weak expected contribution to the EBL from such sources and the lack of sufficiently precise measurements of the EBL density.

The dominant contribution to the EBL density comes from star formation at redshifts of $z\sim1$, where a peak in the SFRD is expected \citep[e.g.][]{hopkins:2006a}. The EBL is, therefore, an excellent probe of the bulk star formation \citep[e.g.][]{dwek:1998a,madau:2000a,chary:2001a}. For example, \citet{fardal:2007a} studied the SFRD by combining the integrated EBL density with measurements of the observed stellar mass density. They  concluded that the measurements for these two quantities are in tension, which could be weakened by adopting a different universal stellar initial mass function (IMF).
{
\citet{horiuchi:2009a} used the integrated EBL density calculated for specific SFRDs together with measurements of the EBL to constrain their modeling of the diffuse neutrino background from supernovae.
}

In this paper, the SFRD will be investigated utilizing limits on the EBL density derived in ME12. In particular, conservative upper limits on the SFRD will be derived. To this end, the EBL density is calculated using emission spectra from simple stellar population (SSP) modeling (Section~\ref{Sec:EBLModel}). The employed methodology is similar to the one presented in \citet{raue:2009a}, but it is extended by making use of the full wavelength dependent EBL constraints and includes modeling of the MIR/FIR emission from dust. The EBL model is constructed to be in minimum tension with the upper limits to derive conservative constraints on the SFRD. The effect of the different parameters entering the model are investigated and upper limits on the SFRD are derived (Section~\ref{Sec:Results}). The implications of the the results are discussed in Section~\ref{Sec:DiscussionConclusions}.

Throughout this work a flat $\Lambda$CDM cosmology with $\Omega_\Lambda=0.7$, $\Omega_\text{M}=0.3$, and $H_0 = 70\,\text{km}\,\text{s}^{-1}\,\text{Mpc}^{-1}$ ('737') is assumed.


\section{EBL model calculations}\label{Sec:EBLModel}

The EBL density is calculated following \citet{dwek:1998a} and \citet{kneiske:2002a} (KN02 in the following) from the star formation rate density ${\rho}_{\ast}$ (SFRD) and the specific luminosity $L_{\nu}(\tau)$ of a simple stellar population (SSP) model of age $\tau$ as, e.g., derived from stellar population syntheses models. The co-moving emissivity (luminosity density) at redshift $z$ is obtained from the convolution
 \begin{equation}
 \mathcal{E}_{\nu}(z) = \int_z^{z_{m}}
 L_{\nu}(t(z)-t(z'))\,\rho_{\ast}(z') \left | \frac{dt'}{dz'} \right |
 dz' ~,
 \label{eq:emissivity}
 \end{equation}
 where the SFRD
 {
 ${\rho}_{\ast}(z)$ is assumed to start at some finite epoch
 $z_m=z(t_m)$ and $t(z)$ is the cosmic time corresponding to a redshift $z$.
 }
 For given evolution of the emissivity a second integration over redshift yields the energy density, or, after multiplication with $\nu c/4\pi$, the co-moving spectral energy distribution (SED) of the EBL
 \begin{equation} P_\nu(z) = \nu
 I_{\nu}(z) = \nu \frac{c}{4\pi} \int_z^{z_m}  \mathcal{E}_{\nu'}(z')
\left | \frac{dt'}{dz'} \right |  dz' ~ ,
\label{eq:ebl} 
\end{equation}
 with $\nu'=\nu(1+z')/(1+z)$.

In the following, the various model parameters are discussed.

\subsection{IMF, metallicity, and SSP model}\label{Sec:SSPIMF}

\begin{figure}
\centering
\includegraphics[width=0.49\textwidth]{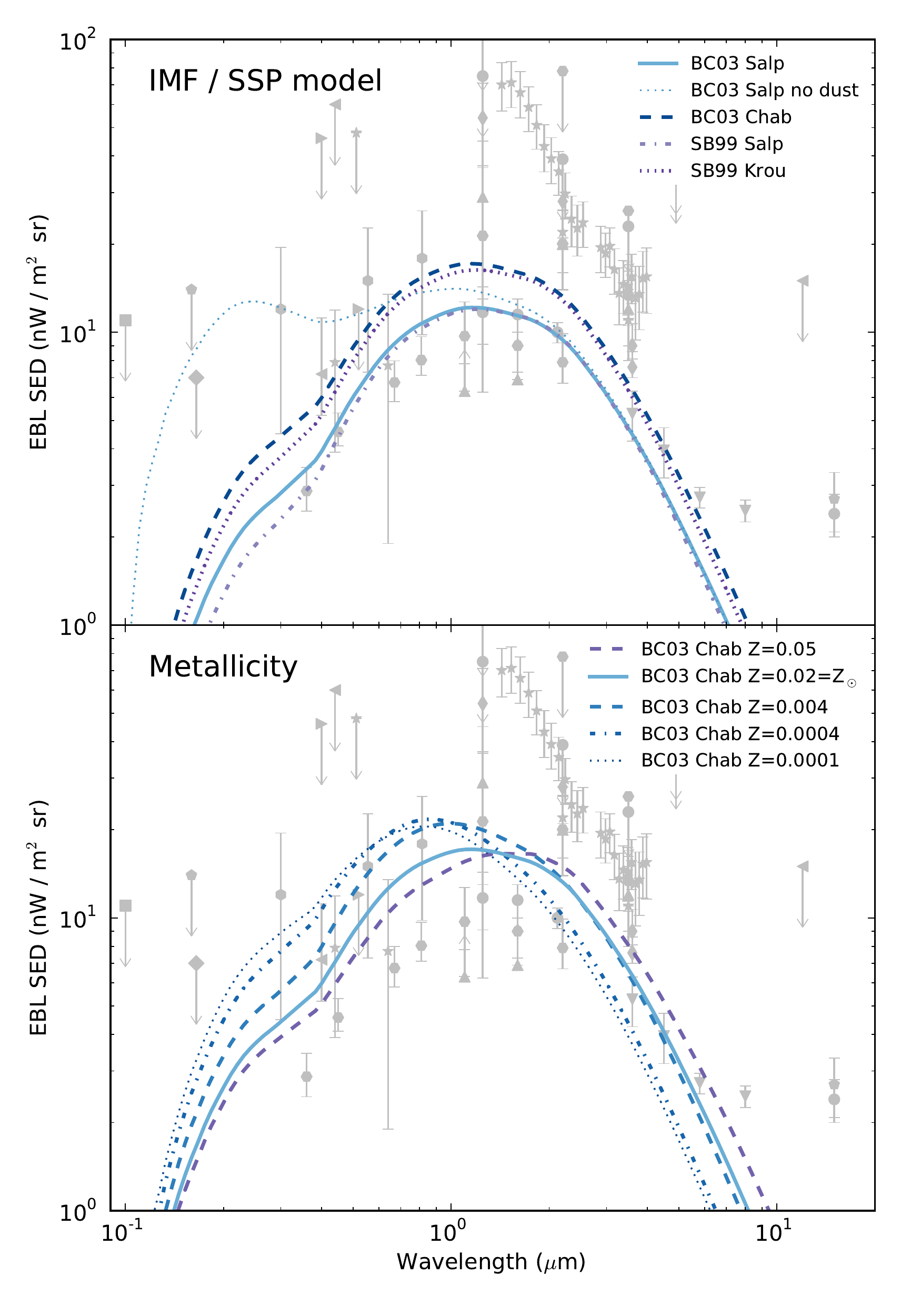}
\caption{Resulting EBL SED ($z=0$) for different choices of parameters of the EBL model. \textit{Upper Panel:} IMF and SSP models. \textit{Lower Panel:} Metallicities.
{
To guide the eye, various EBL measurements and upper and lower limits are shown as grey markers in the background \citep[from][]{hauser:2001a,mazin:2007a,raue:2011a}.
}
}
\label{Fig:EBLModel1}
\end{figure}
 
The SSP spectra used for the EBL modeling are taken from \citet{bruzual:2003a} (BC03 in the following).

\paragraph*{IMF}
The initial mass function (IMF) is one of the fundamental parameters defining the emission properties of the SSP spectra \citep[see, e.g., ][for a recent review on the stellar IMF]{kroupa:2011a}. Experimentally determining the IMF is a challenging tasks. A universal IMF has been found from studies of stellar populations in the Milky Way \citep[Kroupa IMF,][]{kroupa:2001a}, but recent studies of other galaxies indicate that low-mass stars might be more numerous in late-type galaxies \citep{vandokkum:2010a} and that in early-type galaxies the IMF varies systematically with the mass-to-light ratio \citep{treu:2010a,cappellari:2012a}. In particular, \citet{cappellari:2012a} found that the IMF for a large sample of early type galaxies vary from a Kroupa to a Salpeter IMF \citep{salpeter:1955a}, providing more low mass stars ($m<0.5\,\text{M}_\odot$).

BC03 provide SSP model spectra for a Salpeter (Salp) and a Chabrier (Chab) IMF \citep[an IMF very similar to the Kroupa IMF; for a comparison see, e.g., Figure 28 of][]{kroupa:2011a} in the mass range from 0.1 to 100\,$\text{M}_\odot$. The resulting EBL densities from the model calculations for a Salp and a Chab IMF are compared in Figure~\ref{Fig:EBLModel1} Top Panel. The EBL density for a Chab IMF is $\sim30\,\%$ higher then the one resulting from a Salp IMF (all other model parameters are kept the same).
{
This is an effect of the break of the Chab IMF at $\sim0.5\,\text{M}_\odot$: for a mass normalized IMF a Chab IMF produces more stars in the region above $1\,\text{M}_\odot$ compared to the Salp IMF, which does not show this break \citep[see Figure 28 of][]{kroupa:2011a}. Since the bulk of the emission is produced by stars with $>1\,\text{M}_\odot$ adopting a Chab IMF results in more luminosity per mass.
}

A Chab IMF will be adopted for the fiducial model, but results for a Salp IMF will also be presented. While a Chab IMF will result in stronger absolute constraints on the SFRD it turns out that, due to the different behavior of instantaneous and late-time tracers for the SFRD for a specific IMF (see Appendix~\ref{Sec:IMFSalpToChab}), adopting a Chab IMF results in weaker constraints relative to the SFRD derived from instantaneous tracers.

\paragraph*{Metallicity}
The history of the metallicity of stars and star forming regions can be investigated using stars in our galaxy \citep[e.g.][]{edvardsson:1993a} or large samples of distant galaxies \citep[e.g.][]{panter:2008a}. These studies show that the solar metallicity value of  $Z_\odot=0.02$ is close to the average metallicity over time up to redshifts of at least $z=2$. The metallicity evolution is a function of galaxy mass: while massive galaxies ($M>10^{11}\,M_\odot$) are enriched early and show an almost constant metallicity since redshift of $z\sim2$, lower mass galaxies show a decrease in metallicity \citep{panter:2008a}. 
{
This decrease in metallicity can also be seen in the stars of our galaxy, where old stellar populations show lower metallicities \citep[e.g.][]{edvardsson:1993a} while slightly higher metallicities then the solar metallicity (up to [Fe/H]$\sim0.2$) are observed in present day star forming clouds \citep{binney:1998a}. Overall, the average metallicity at redshifts greater than $0.2$ is well described by a solar metallicity or less \citep{panter:2008a}.
}

The impact of different metallicities on the resulting EBL density is displayed in Figure~\ref{Fig:EBLModel1} bottom panel. A sub-solar metallicity value leads to an increased emission in the UV to NIR up to wavelengths of $\sim1-2\,\mu{}m$ due to the lower mass loss in massive stars and hence increased time-integrated emissivity. A metallicity higher then the solar value leads to an increased EBL density in the NIR but less emission in the UV to O. A solar metallicity is adopted as the fiducial value for the EBL modeling and the effect of lower and higher metallicities on the resulting constraints on the SFRD will be investigated.

\paragraph*{SSP model}
For comparison, Figure~\ref{Fig:EBLModel1} Top Panel also shows the resulting EBL SED adopting SSP spectra calculated with the Starburst99 code \citep[Version 6.0.3;][]{leitherer:1999a,vazques:2005a,leitherer:2010a}, for a Salp and a Kroupa IMF. Excellent agreement is found with the calculations using the BC03 SSP spectra, with only small difference in the UV.
{
These could arise from the different treatment of, e.g., the thermally pulsing asymptotic giant-branch (TP-AGB) phase (see, e.g., \citealt{bruzual:2003a} for a more detailed comparison of the models). An  increase of the overall luminosity in the TP-AGB phase as, e.g., predicted by \citet{maraston:2005a}, would increase the resulting EBL and strengthen the limits.
}

\subsection{Dust absorption and emission}

\begin{figure}
\centering
\includegraphics[width=0.49\textwidth]{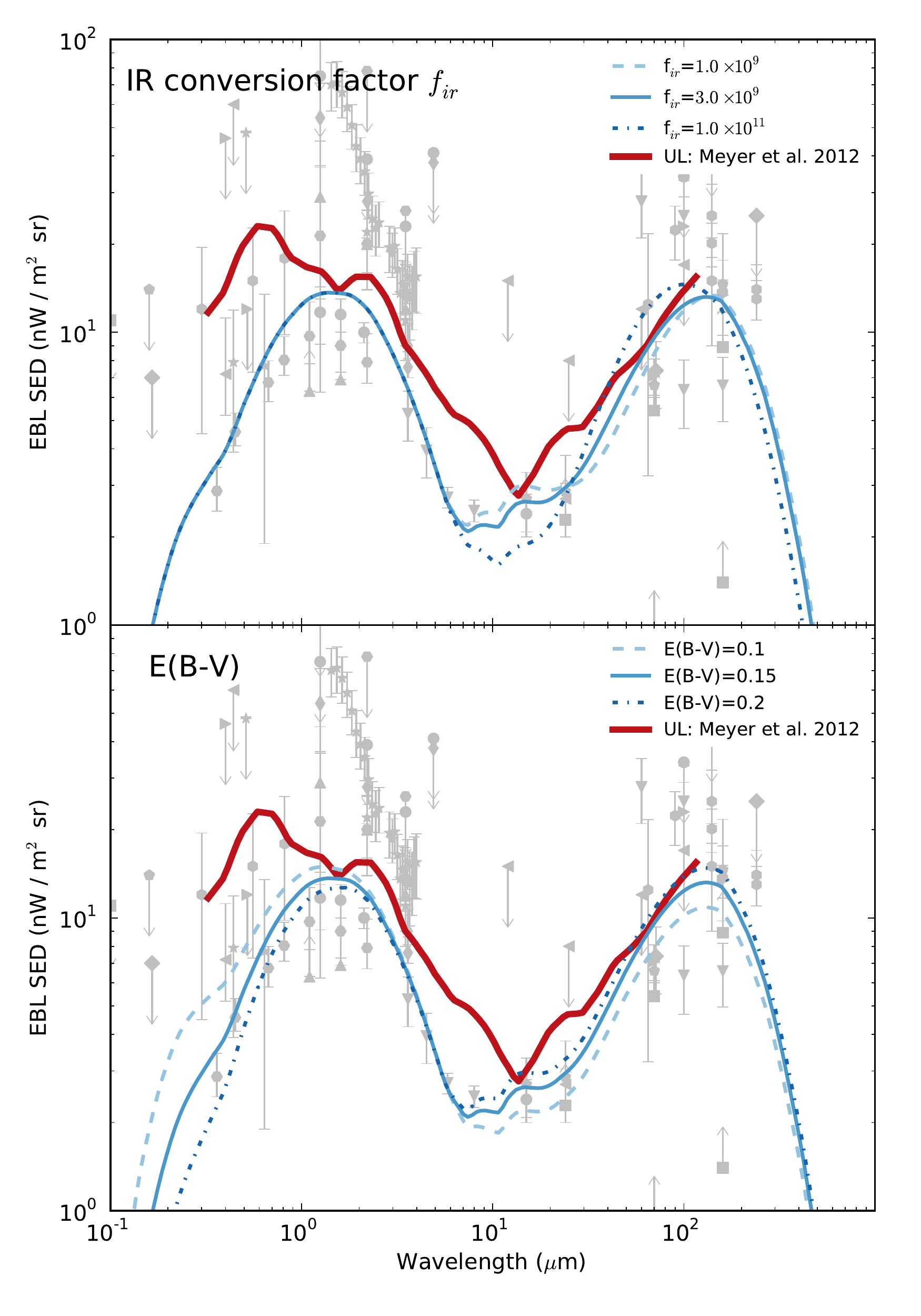}
\caption{Resulting EBL SED ($z=0$) for different choices of the infrared conversion factor $f_{ir}$ (upper panel) and the color of the dust attenuation $E(B-V)$ (lower panel) in comparison the EBL upper limits from ME12 (red thick line).
Shown are results for a Chab IMF, $E(B-V)=0.15$, $f_{ir}=3\times10^9$, and a SFRD with $\beta=0$ and $z_0=1$ if not denoted otherwise in the legend.
{
To guide the eye, various EBL measurements and upper and lower limits are shown as grey markers in the background \citep[from][]{hauser:2001a,mazin:2007a,raue:2011a}.
}
}
\label{Fig:EBLModel2}
\end{figure}

Part of the stellar emission is reprocessed by dust. Dust attenuation is calculated following KN02 and applied directly to the SSP spectra. Full absorption of all ionizing photons is assumed with 50\%\footnote{KN02 used a value of 68\%. Here, conservatively, 50\% is assumed, i.e., less of the ionizing radiation is reprocessed into the EBL.} of the absorbed photons being reemitted as Lyman-$\alpha$ emission, which is consequently absorbed by dust. In addition, an average extinction curve is applied:
\begin{equation}
A_\lambda = 0.68 \cdot E(B-V) \cdot R \cdot (\lambda^{-1} - 0.35)
\end{equation}
with $R=3.2$, $\lambda$ being the wavelength in $\mu{}\text{m}$, and the absorption coefficient being given by $g(\lambda) = 10^{-0.4\cdot{}A_\lambda}$.
KN02 assumed two different $E(B-V)$ values for the young and the old stellar population. Here, for simplicity, a single attenuation curve  for all stellar ages specified by one $E(B-V)$ value is used. The resulting attenuation is in reasonable agreement with more detailed attenuation curves, e.g., the attenuation curves derived by \citet{pei:1992a} for nearby galaxies. Dust attenuation is strongest in the UV to O where the EBL limits considered here are not strongly constraining, and, therefore, the exact shape of the energy dependent dust attenuation is not important for the analysis presented here. Figure~\ref{Fig:EBLModel1} top panel illustrates the effect of dust attenuation on the resulting EBL SED (dotted vs. solid line).

The attenuated photons are reemitted in the MIR and FIR.
The dust reemission is modeled using the galaxy IR SED templates from \citet{chary:2001a} covering a range of IR luminosity from $10^8$ to $10^{13}\,\text{L}_\odot$. The IR SED depends on the total IR luminosity.
To map the galaxy IR SED templates to the SSP spectra, the total dust attenuated luminosity $L^{SSP}_{dust}$ from the SSP spectra is scaled by a factor $f_{ir}$ (in $M_\odot$) and matched to a corresponding integrated galaxy IR luminosity $L^{GAL}_{IR}$ from the IR SED templates:
\begin{equation}
L^{GAL}_{IR} = f_{ir} \times L^{SSP}_{dust} \; .
\end{equation}
The corresponding galaxy IR template is then scaled by a factor $1/f_{ir}$ and added to the SSP spectrum. For values of $f_{ir} \times L^{SSP}_{dust}$ exceeding the range provided by \citet{chary:2001a} the highest/lowest galaxy IR template is scaled accordingly.
{
\footnote{In the following, values of $ f_{ir}$ will be quoted without explicitly noting the unit.}
}

 $f_{ir}$ is a free parameter of the model. Given that the SSP spectra are normalized to the mass of the stellar population, $f_{ir}$ should be of order of a galaxy mass (i.e. $\sim 10^{10} M_\odot$; see \citealt{kneiske:2002a}.). The effect of different choices for $f_{ir}$ on the resulting EBL SED are illustrated in Figure~\ref{Fig:EBLModel2}: increasing values of $f_{ir}$ increase the ratio between the the FIR and MIR EBL.

The parameters $E(B-V)$ and $f_{ir}$ can, therefore, be used to change the overall shape of the EBL SED, i.e., to shift photons from the UV/O/NIR to the MIR or FIR. Since the aim of the paper is to produce conservative upper limits, $E(B-V)$ and $f_{ir}$ are used to construct an EBL model which has \emph{minimal tension} with the EBL limits used in the analysis, i.e., which produces an EBL SED following closely the limit. This will ensure conservative upper limits.

The resulting EBL SED for the optimum choice of parameters, $E(B-V)=0.15$ and $f_{ir}=3\times10^9$, is shown in Figure~\ref{Fig:EBLModel2}. It follows very closely the upper limits from ME12. For smaller or larger values of $f_{ir}$ the resulting EBL SED exceeds the EBL upper limits in the MIR and FIR, respectively (Figure~\ref{Fig:EBLModel2} Upper Panel). A higher/lower $E(B-V)$ leads to overall more/less attenuation and, thereby, decreases/increases the UV/O/NIR relative to the MIR and FIR emission (Figure~\ref{Fig:EBLModel2} Lower Panel). The effect of different choices of $E(B-V)$ on the limits will be discussed in Section~\ref{Sec:Results}.

\subsection{Star formation rate density}

\begin{figure}
\centering
\includegraphics[width=0.5\textwidth]{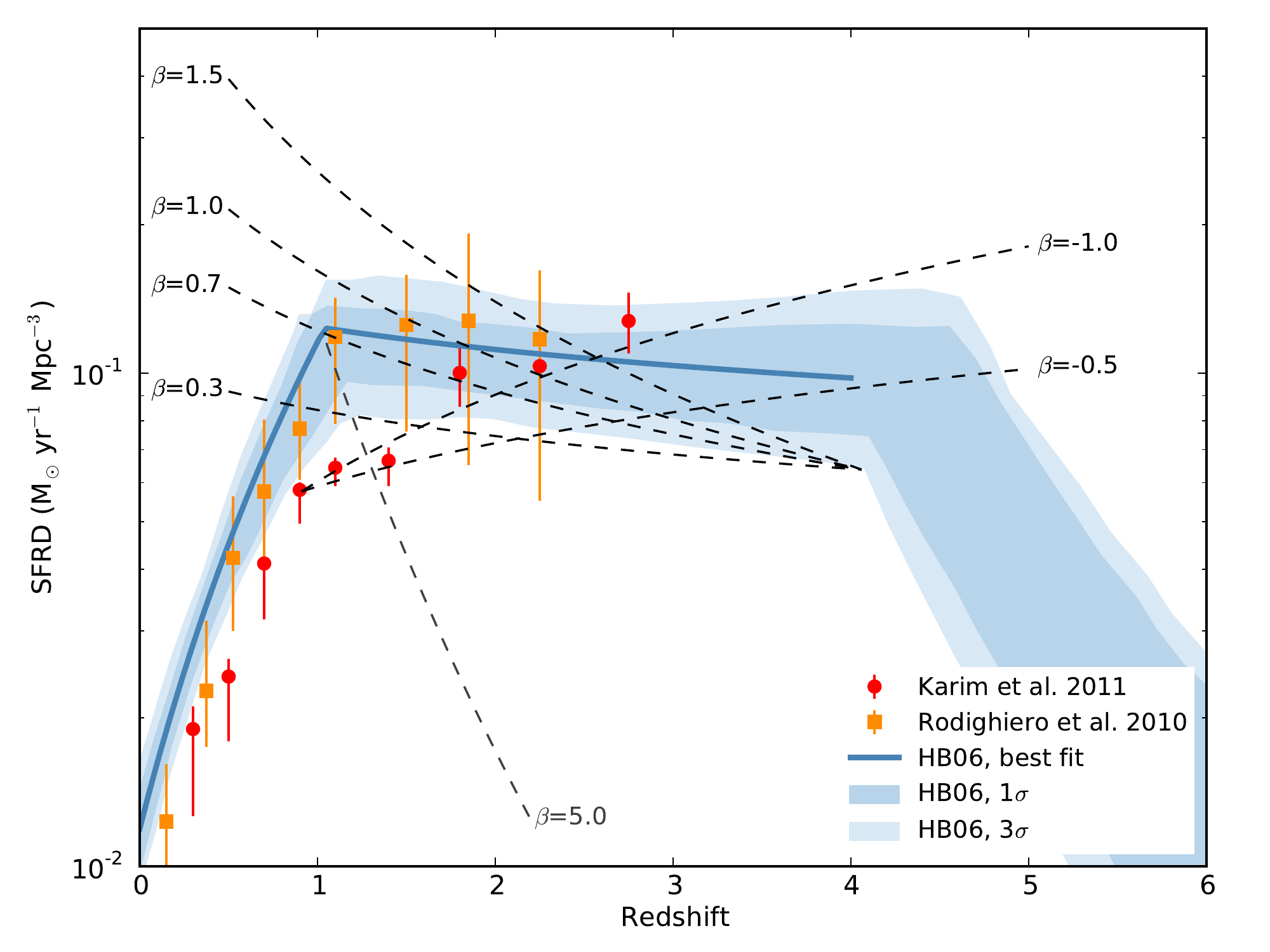}
\caption{Star formation rate density (SFRD) vs redshift. Shown are the best fit broken power law function and the 1 and $3\,\sigma$ contours from HB06 and measurements from \citet{rodighiero:2010a} and \citet{karim:2011a}. The dashed and dotted lines illustrate the behavior of a power law function in $z+1$ for different choices for the power law index $\beta$. For $\beta > 0$ the functions are normalized at $z=4$ at the lower edge of the  $3\,\sigma$ contour from HB06,  for $\beta < 0$ the functions are normalized at $z=0.9$ to the measurement from \citet{karim:2011a}. All values have been converted to a Chab IMF using a scaling factor of 1.65 (see Appendix \ref{Sec:IMFSalpToChab} for details).}
\label{Fig:SFRD}
\end{figure}

The SFRD, as derived from a large compilation of different measurements by \citet[HB06 in the following]{hopkins:2006a}, together with two recent estimates by \citet[RO10 in the following]{rodighiero:2010a} (IR) and \citet[KA11 in the following]{karim:2011a} (radio) is shown in Figure~\ref{Fig:SFRD}. The SFRD up to redshifts of $z\sim4$ is characterized by a steep rise up to redshifts of $z\sim1$ followed by a flatter behavior up to redshifts of $z=3-4$. While the data agree on the overall shape, two different cases can be distinguished, exemplarily represented by the RO10 and the KA11 data: in the former case, a higher peak normalization of the SFRD is expected, followed by a flat or weakly declining SFRD for $z>1$. In the latter case, the peak SFRD is lower ($O(0.2\,\text{dex})$) but then is followed by a still rising SFRD with a shallower slope up to redshift $z\sim3-4$ \citep[a trend also found in H-$\alpha$ data by][]{sobral:2012a}. At higher redshift the SFRD is less certain and different tracers produces different results: while observations of Lyman break galaxies in deep Hubble Space Telescope observations indicate a rapid decline for $z>4$ \citep[e.g.][]{bouwens:2009a} SFRDs derived from gamma-ray burst observations indicate a flat behavior \citep{yuksel:2008a,kistler:2009a}. In this study it will be focused on the SFRD up to redshifts of $z=4$, in particular on the peak of the SFRD around $z\sim1$.

Overall, the SFRD $\rho_{\ast}(z)$ up to a redshift of $z=4$ can be well described by a broken power law in $z + 1$:
\begin{equation}
\rho_{\ast}(z) = \rho_0 \left(\dfrac{z+1}{z_0 + 1}\right)^{-\Gamma}
\end{equation}
with
\begin{equation}
\Gamma = \begin{cases}
\; \alpha  & \text{for $z \leqslant z_0$} \\
\; \beta & \text{for $z > z_0$} \\
\end{cases} \; .
\end{equation}
$z_0$ being the redshift of the peak and $\rho_0$ being the SFRD normalization at $z_0$.

This broken power law description of the SFRD is adopted for the EBL modeling. $\rho_{\ast}(0)$ is fixed to a value derived from the best fit broken power law from HB06: $\rho_{\ast}(0) = 0.02$ and $0.012$ for the Salp and the Chab IMF, respectively. Varying $\rho_{\ast}(0)$ in the $3\,\sigma$ errors of HB06 ($O(30\,\%)$) does only result in small changes of the resulting EBL density ($O(1-7\,\%)$, depending on the wavelength), mainly in the UV (and then, through dust attenuation and reemission in the MIR/FIR). $z_0$, $\rho_0$, and $\beta$ are chosen as free parameters, which will be varied, while $\alpha$ is given by
\begin{equation}
\alpha = \frac{\log \rho_0 - \log \rho_{\ast}(0)}{\log (z_0 + 1)} \; .
\end{equation}

{
The different values of $\beta$ adopted in this study are illustrated in Figure~\ref{Fig:SFRD}. Two different cases can be identified: a rising ($\beta<0$) and a falling ($\beta>0$) second slope of the SFRD. For the former case, i.e., an increasing SFRD at $z>z_0$, $\beta=-1$ with a maximum redshift of $z_{max}=3$ is adopted, which is a good representation of the SFRD found by KA11 for $z>0.9$. In addition, $\beta=-0.5$ is used, which results in a reasonable representation of the SFRD for increased values of $\rho_0$ at $z\sim1$ above the value from KA11. For the latter case, i.e., a decreasing SFRD at $z>z_0$, the choice for the different $\beta$s is derived from the $3\,\sigma$ contour of the broken power law fit to SFRD measurements from instantaneous tracers by HB06 (Figure~\ref{Fig:SFRD}): the lower edge of the $3\,\sigma$ contour at $z=4$ is used as normalization for the SFRD to derive different slopes which intersect the contours at $z=1$ at the minimum ($\beta=0.3$), the best fit ($\beta=0.7$), and the maximum ($\beta=1$) allowed value. In addition, $\beta=1.5$ is used which connects the minimum value at $z=4$ with the maximum value at $z=2$. As an extreme case not compatible with the SFRD values derived from instantaneous tracers  $\beta=5$ is considered, corresponding to a sharp drop of the SFRD after the peak.
}


\section{Method}\label{Sec:Method}

\begin{figure}
\centering
\includegraphics[width=.49\textwidth]{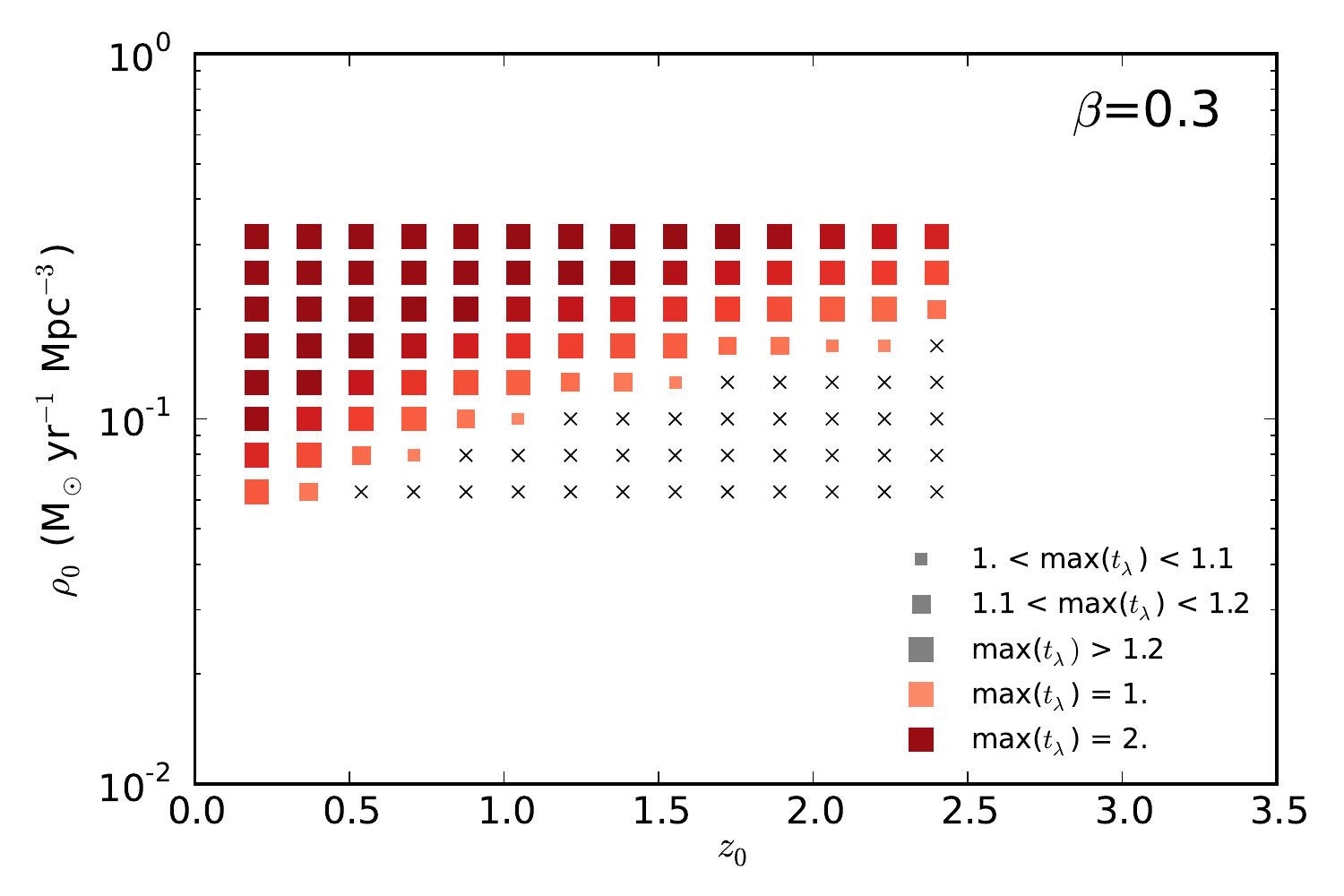}
\caption{Limits on the SFRD parameters $z_0$ and $\rho_0$  (i.e. the peak redshift and normalization) of the EBL modeling for $\beta=0.3$. The other model parameters are set to the fiducial values (Chab IMF, $\rho_{\ast}(0)=0.012\,\text{M}_\odot\,\text{yr}^{-1}\,\text{Mpc}^{-3}$, $E(B-V)=0.15$, and $f_{ir}=3\times10^{9}$). Squares and crosses denote the values of $z_0$ and $\rho_0$ for specific EBL models tested. $z_0$-$\rho_0$-pairs shown with a square symbol result in an EBL density in conflict with the upper limits (see legend in the lower right panel for details).
{
}
}
\label{Fig:ResultsBeta}
\end{figure}

A grid in $z_0$ vs $\rho_0$ is constructed to investigate different SFRDs in the EBL modeling, covering a range of $z=0.2$ to $z=2.4$ in redshift and $\log(\rho_\ast) = -1.2$ to $\log(\rho_\ast) = -0.5$ in peak SFRD (see Figure~\ref{Fig:ResultsBeta}). For each point $(z_0, \rho_0)$ of the grid the EBL density is calculated. The resulting EBL SED  is then compared with the EBL limits from ME12 by calculating the ratio between the EBL SED $P_\lambda(0)$ and the limits $L_\lambda$ at different wavelengths:
\begin{equation}
t_\lambda = \frac{P_\lambda(0)}{L_\lambda}
\end{equation}

 If the ratio $t_\lambda$ exceeds $1$ at any wavelength, the EBL model parameters are considered to be in conflict with the limit.

Two different cases can be discerned: (i) if the maximum ratio value exceeds one (i.e. $\text{max}(t_\lambda) > 1$) the resulting EBL is considered in \emph{weak} tension with the limits. (ii) Allowing for a systematic error of the EBL limit of 20\% (see Appendix~\ref{Sec:SysErrEBLLim} for details), the EBL is considered to be in \emph{strong} tension with the limits if the maximum ratio value at any wavelength  exceed $1.2$ ($\text{max}(t_\lambda) > 1.2$).

{
Figure~\ref{Fig:ResultsBeta} shows the results for a Chab IMF for $\beta = 0.3$.
}
The squares and crosses denote the value of $z_0$ and $\rho_0$ (i.e. the peak redshift and normalization) for specific EBL model tested. The marker symbols and colors give the ratio of EBL SED vs limits (see the legend in the lower right panel of the Figure). Small squares represent \emph{weak} tension with the upper limit, while big squares represent \emph{strong} tension.

{
The constraints on $z_0$ and $\rho_0$ are converted into constraints on the SFRD as follows: for each $z_0$ value tested the $\rho_0^\text{W/S}$ value is calculated for which $\text{max}(t_\lambda) = 1/1.2$, i.e., for which the \emph{weak/strong} tension limiting value is reached. This is done via linear interpolation between the discrete values in $t_\lambda$ vs $\rho_0$. For each value pair $(z_0, \rho_0^\text{W/S})$ the SFRD function $\rho(z, z_0, \rho_0^\text{W/S})$ is evaluated. The limit on the SFRD for a specific redshift $z_i$ is then taken to be the maximum of the different $\rho(z_i, z_0, \rho_0^\text{W/S})$ functions, i.e., an upper envelope is calculated.
}


\section{Results}\label{Sec:Results}

\begin{figure}
\centering
\includegraphics[width=.49\textwidth]{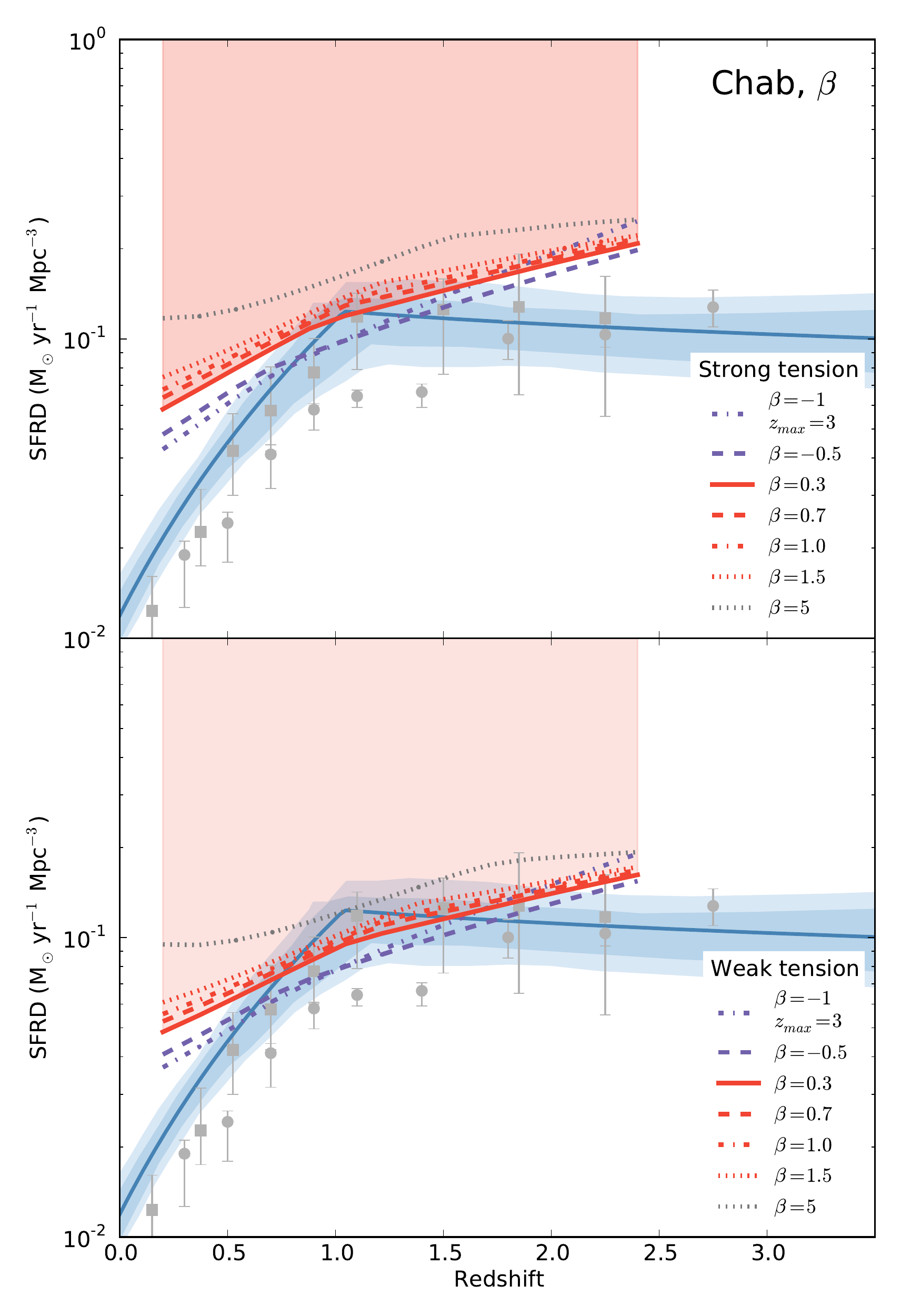}
\caption{Limits on the SFRD for different values of $\beta$. Other EBL model parameters are fixed to the fiducial values (Chab IMF, $\rho_{\ast}(0)=0.012\,\text{M}_\odot\,\text{yr}^{-1}\,\text{Mpc}^{-3}$, $E(B-V)=0.15$, and $f_{ir}=3\times10^{9}$). The light red filled area corresponds to the exclusion region for the fiducial value $\beta=0.3$. The blue line and filled areas are the best fit and the 1 and $3\,\sigma$ contours from HB06, respectively. Grey markers are the measurements from RO10 and KA11 (see Figure~\ref{Fig:SFRD} for details).}
\label{Fig:ResultsBetaSummary}
\end{figure}
 
\paragraph*{SFRD $\bm\rho_\ast$ / $\bm\beta$}
The upper limits on the SFRD for different values of $\beta$ are summarized in Figure~\ref{Fig:ResultsBetaSummary}.
First, it can be noted that for all choices of $\beta$
{
compatible with the SFRD values from instantaneous tracers
}
the SFRD is constrained to $\lesssim 0.1\,\text{M}_\odot\,\text{yr}^{-1}\,\text{Mpc}^{-3}$ and $< 0.2\,\text{M}_\odot\,\text{yr}^{-1}\,\text{Mpc}^{-3}$ for $z\sim1$ and $z\sim2$, respectively. For $z\sim1$ this value is below the best fit value from HB06 for their broken power law fit. For $\beta \leqslant0.3$ the $1\,\sigma$ range from HB06 is in weak tension with the limits.
The measurements from KA11 are not in conflict with the upper limits derived here, even for a rising SFRD ($\beta<0$) beyond the peak redshift $z_0$. In the following, $\beta=0.3$ will be adopted as the fiducial value, which is also the best fit value found by \citet{yuksel:2008a} in their broken power law fit of a recent SFRD data compilation including data from gamma-ray burst.
{
Higher values of $\beta$ ($>1.5$), which are in conflict with the values derived from instantaneous tracers (Figure~\ref{Fig:SFRD}), would lead to weaker limits. But even for the extreme case of $\beta=5$, corresponding to an almost instantaneous drop of the SFRD behind the peak, the limits are still in weak tension with the HB06 values.
}

\begin{figure}
\centering
\includegraphics[width=.49\textwidth]{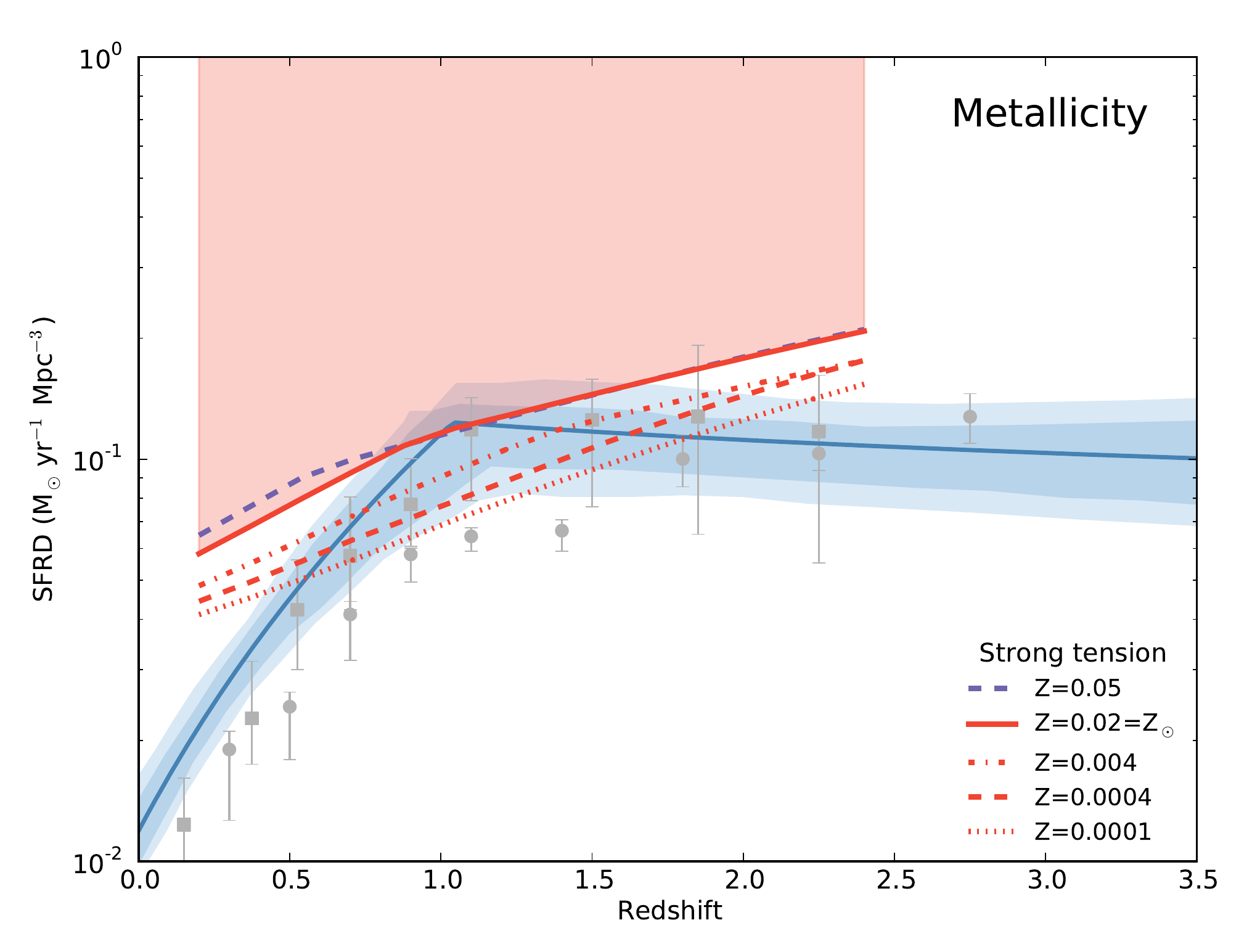}
\caption{Limits on the SFRD for different metallicities of the SSP spectra. Other EBL model parameters are fixed to the fiducial values (Chab IMF, $\rho_{\ast}(0)=0.012\,\text{M}_\odot\,\text{yr}^{-1}\,\text{Mpc}^{-3}$, $\beta=0.3$, $E(B-V)=0.15$, and $f_{ir}=3\times10^{9}$). The light red filled area corresponds to the exclusion region for the fiducial metallicity value $\text{Z}=0.02=\text{Z}_\odot$. Other markers are as in Figure~\ref{Fig:ResultsBetaSummary}.}
\label{Fig:ResultsMetallicity}
\end{figure}

\paragraph*{Metallicity}
Figure~\ref{Fig:ResultsMetallicity} displays the resulting upper limits on the SFRD for different metallicities of the SSP. With decreasing metallicity the constraints get significantly stronger, with the limit being $\sim\text{0.2\,dex}$ lower for $\text{Z}=10^{-4}$ in comparison to the limit for the fiducial metallicity value of $\text{Z}=0.02=\text{Z}_\odot$. This is a direct result of the increased UV, O, and NIR emission for the sub-solar metallicity models (Figure~\ref{Fig:EBLModel1} bottom panel). Since the metallicity increased over the history of the universe, i.e., higher redshift should have lower metallicities, adopting solar metallicity as fiducial value is a conservative choice. For a metallicity higher then the solar value the limits are very similar to the solar metallicity ones, with only small difference at at low redshift $z<0.75$, where the UV emission from young stars provides a larger contribution to the total EBL.

\begin{figure}
\centering
\includegraphics[width=.49\textwidth]{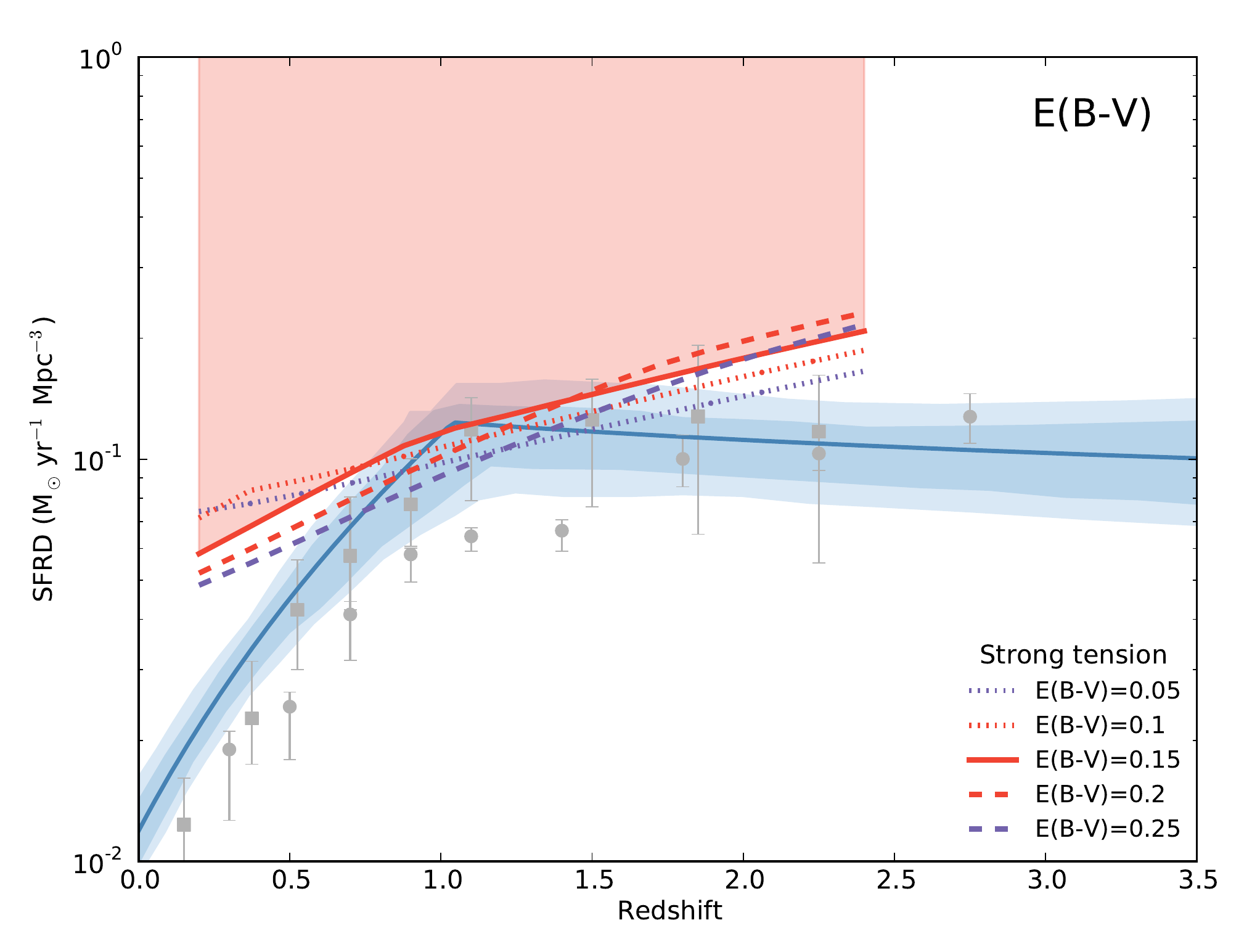}
\caption{Limits on the SFRD for different strengths of the dust attenuation parameter $E(B-V)$. Other EBL model parameters are fixed to the fiducial values (Chab IMF, $\rho_{\ast}(0)=0.012\,\text{M}_\odot\,\text{yr}^{-1}\,\text{Mpc}^{-3}$,  $\beta=0.3$, and $f_{ir}=3\times10^{9}$). The light red filled area corresponds to the exclusion region for the fiducial dust attenuation value $E(B-V)=0.15$. Other markers are as in Figure~\ref{Fig:ResultsBetaSummary}.}
\label{Fig:ResultsEBV}
\end{figure}

\paragraph*{Dust absorption}
The effect of different strengths of the dust attenuation parameter $E(B-V)$ on the resulting SFRD limits is displayed in Figure~\ref{Fig:ResultsEBV}. Varying $E(B-V)$ from the fiducial value by $\pm0.05$ does not strongly change the results: A larger value $E(B-V)=0.2$ leads to stronger constraints in the redshift range up to $z\sim1.5$ and marginally relaxed limits at higher redshift. Lowering the value to $E(B-V)=0.1$ results in stronger constraints at redshift $z\gtrsim0.5$ and marginally weaker limits below. This behavior can be understood from the correlation of the peak position of the EBL SED with the peak of the SFRD: a low-redshift peak of the SFRD leads to lower-wavelength UV/O peak of the EBL SED. While in the MIR/FIR the rising (low-wavelength) flank of the EBL SED peak is strongly constrained by the EBL limits the UV/O EBL is not strongly constrained. A model with a lower $E(B-V)$ value produces less MIR/FIR EBL and more UV/O, which leads to a relaxation of the limits for lower-redshift-peaked SFRD. A similar argument can be made for stronger dust attenuation and higher SFRD peak redshifts. For $E(B-V)$ values $\pm0.1$ from the fiducial value the same trend are visible but the limit overall get stronger.

\begin{figure}
\centering
\includegraphics[width=.49\textwidth]{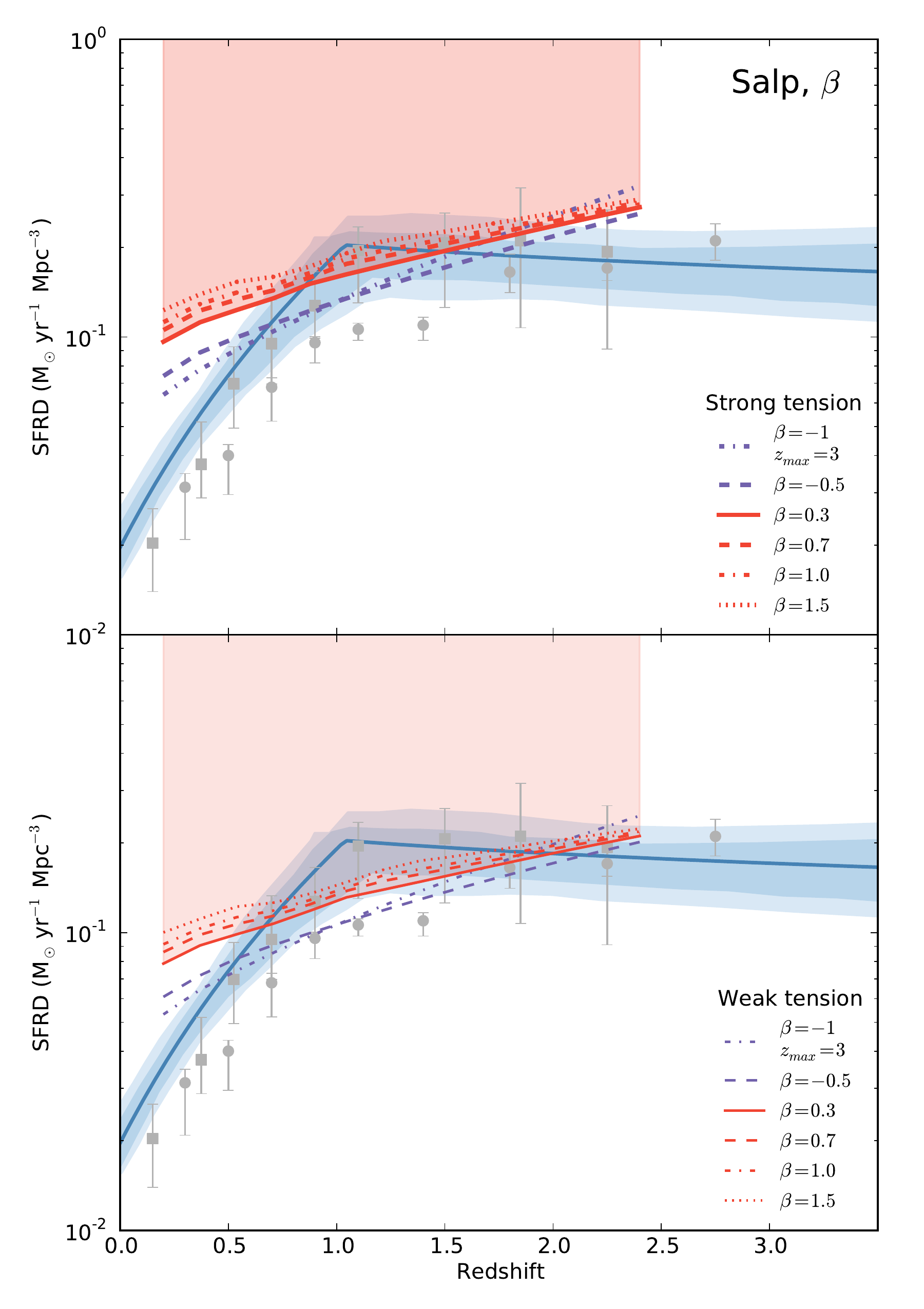}
\caption{Limits on the SFRD for a Salp IMF for different values of $\beta$. Other EBL model parameters are fixed to fiducial values ($\rho_{\ast}(0)=0.02\,\text{M}_\odot\,\text{yr}^{-1}\,\text{Mpc}^{-3}$, $E(B-V)=0.15$, and $f_{ir}=3\times10^{9}$). The light red filled area corresponds to the exclusion region for the fiducial value $\beta=0.3$. Other markers are as in Figure~\ref{Fig:ResultsBetaSummary}.}
\label{Fig:ResultsIMF}
\end{figure}
 
 \paragraph*{Initial mass function}
As discussed in Section~\ref{Sec:SSPIMF} a Chab IMF has been adopted as fiducial value for the EBL modeling, but indications for variations of the IMF do exist.
In Figure~\ref{Fig:ResultsIMF} the SFRD limits for an EBL model with an Salp IMF are shown for different choices of $\beta$. In comparison to the SFRD derived from instantaneous tracers the limits are more constraining then in the case of a Chab IMF effectively excluding the full $3\,\sigma$ range from HB06 at $z\sim1$.

\begin{figure}
\centering
\includegraphics[width=.49\textwidth]{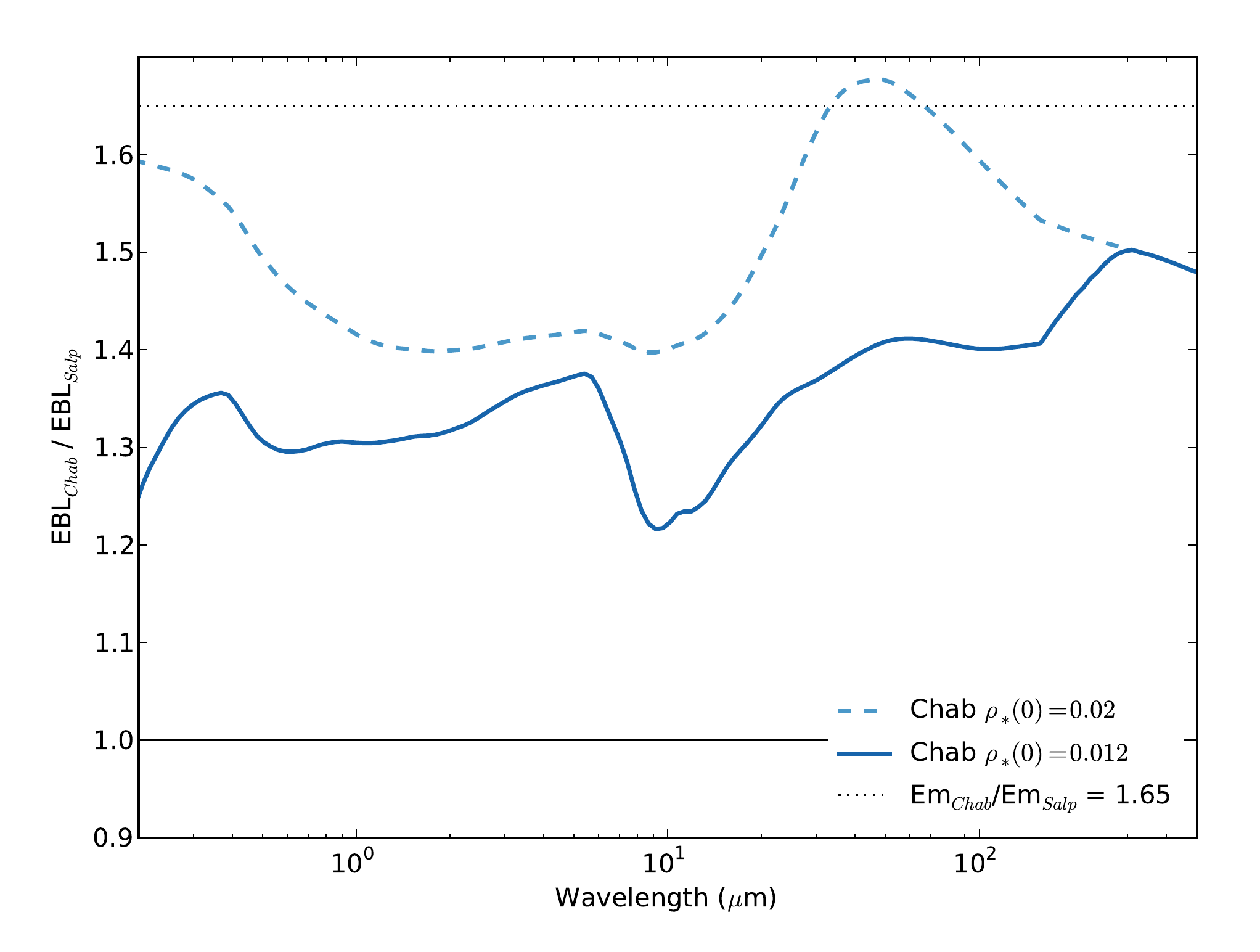}
\caption{Ratio of the EBL SED for a Chab and Salp IMF. Shown are results for two different values of $\rho_{\ast}(0)$ for the Chab IMF: $0.012\,\text{M}_\odot\,\text{yr}^{-1}\,\text{Mpc}^{-3}$, the fiducial value, and $0.02\,\text{M}_\odot\,\text{yr}^{-1}\,\text{Mpc}^{-3}$, the Salp IMF value.}
\label{Fig:ResultsEBLComparison}
\end{figure}
 
This difference can be understood from the time evolution of the ratio of the SSP emissivities for a Salp and a Chab IMF: while for young ages of the SSP  $<10^7$\,yrs the ratio for the luminosities between Chab and Salp IMF is $f_\text{IMF} \sim1.65$ it decreases for larger ages (see Figure~\ref{Fig:IMFSalpToChab1} and discussion in Appendix~\ref{Sec:IMFSalpToChab}). The EBL effectively samples a large time range of order of the age of the universe. Since the SFRD displays a peak at a redshift of $z\sim1$, the EBL at $z=0$ is already influenced by the emission from older stars, resulting in a decreased ratio for the EBL \citep[see also][Section C]{horiuchi:2009a}. This effect can also be seen in Figure~\ref{Fig:ResultsEBLComparison}, where the ratio between EBL SEDs calculated with a Chab and a Salp IMF are shown. For typical model assumption for the SFRD the ratio in the EBL SED is between 1.3 and 1.4 in the wavelength range of $1-10\,\mu{}m$, i.e., much lower then the 1.65 used as IMF scaling factor for the SFRD from instantaneous tracers.  \citet{horiuchi:2009a}.
 
\begin{figure}
\centering
\includegraphics[width=.49\textwidth]{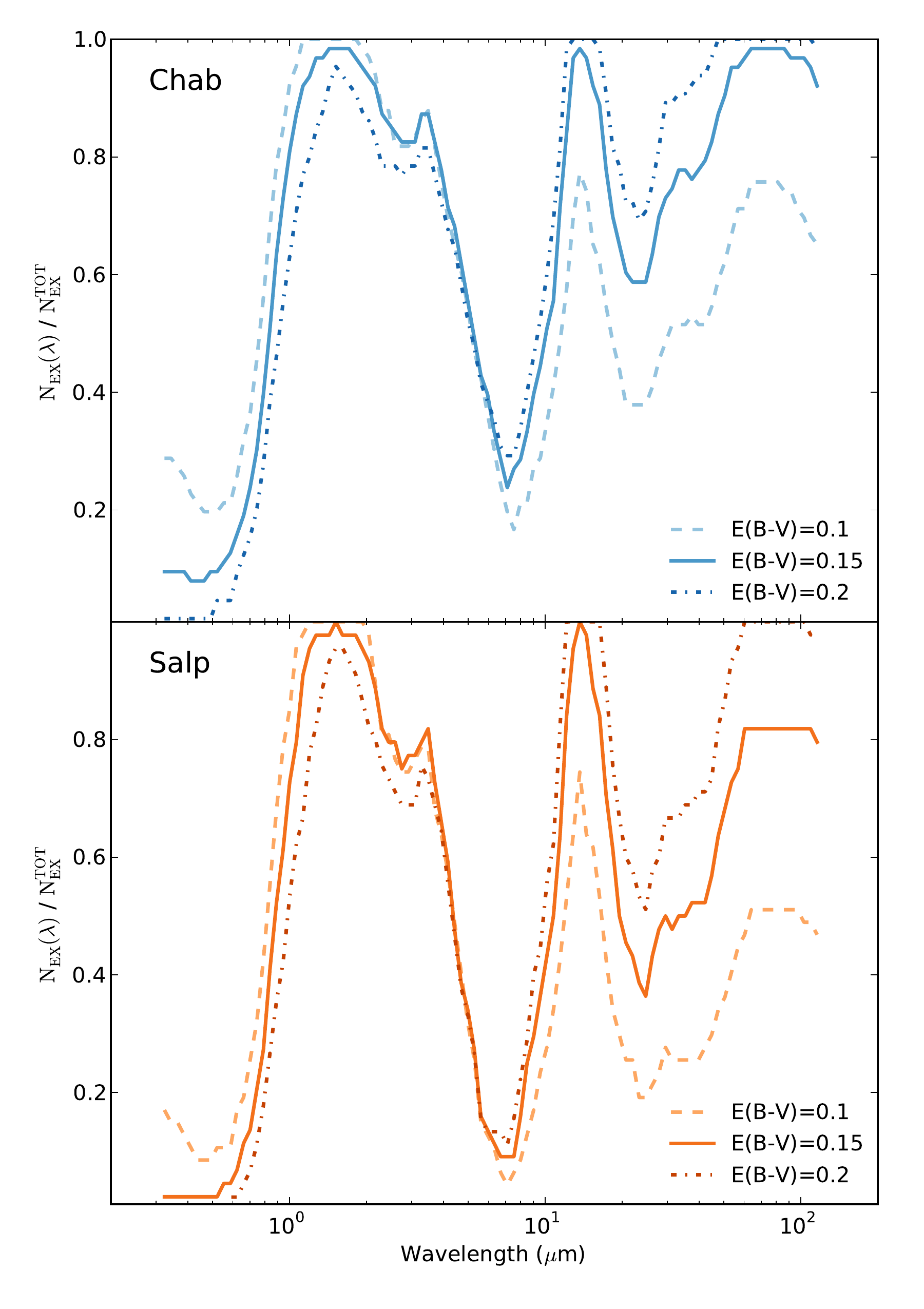}
\caption{Number of EBL models excluded at a certain wavelength $\text{N}_\text{EX}(\lambda)$ divided by the total number of excluded models  $\text{N}_\text{EX}^\text{TOT}$. In the upper panel a Chab IMF in the lower panel a Salp IMF is used for the EBL modeling, respectively. The $z_0/\rho_0$ grid from Figure~\ref{Fig:ResultsBeta} is adopted and different values of $E(B-V)$ are used. Other EBL model values are kept to the fiducial values ($\rho_{\ast}(0)=0.012/0.02\,\text{M}_\odot\,\text{yr}^{-1}\,\text{Mpc}^{-3}$ for a Salp/Chab IMF, $\beta=0.3$, and $f_{ir}=3\times10^{9}$).}
\label{Fig:ResultsStats}
\end{figure}

\paragraph*{EBL wavelength dependency} 

{
Since the EBL limits cover a range of over two decades in wavelength it is interesting to investigate at which wavelength the tension between the limits and the modeling actually occur. Several parameters entering the EBL modeling have no or only minor impact on the resulting EBL SED shape: changing $\rho_0$ changes the overall normalization of the resulting EBL (see Equation~\ref{eq:emissivity} and \ref{eq:ebl}). The effect of varying $z_0$ on the EBL SED shape is more complicated, mainly resulting in a shift of the EBL SED due to the increased redshift of the epoch with the maximum emission. Given the limited range for $z_0$ considered here ($\sim0.25-2.25$, i.e., a maximum shift in wavelength by a factor of three) the effect is also not overly strong but should be kept in mind.

To investigate the EBL wavelength dependency the relative rejection at a specific EBL wavelength is calculated using a set of models marginalized over $\rho_0$ and $z_0$. The number of EBL models excluded at a certain wavelength $\text{N}_\text{EX}(\lambda)$ is divided by the total number of excluded models $\text{N}_\text{EX}^\text{TOT}$. Only models in \emph{strong} tension are considered (though the results do not significantly change when also including models with \emph{weak} tension).
The results of this analysis are shown in Figure~\ref{Fig:ResultsStats} for a Chab (upper panel) and a Salp IMF (lower panel) and different values of $E(B-V)$.
}

It can be seen from the upper panel that for the fiducial choice of parameters (i.e. $E(B-V)=0.15$) a larger fraction ($\sim98\,\%$) of the shapes are excluded in the NIR, MIR, and FIR simultaneously, i.e., the methods presented here take full advantage of the wide wavelength range of the EBL upper limits. This behavior is expected from the construction of the model: the NIR/MIR/FIR ratios have been tuned to closely follow the EBL upper limits. A lower value of $E(B-V)$ leads to more exclusion in the NIR, a higher value to an increased exclusion in the FIR. The results using a Salp IMF are very similar, though a slightly higher value of $E(B-V)$ somewhere between 0.15 and 0.2 should lead to a more homogeneous exclusion over the wavelength range. The simultaneous tension in the three wavelength regimes (NIR/MIR/FIR) make the SFRD limits robust against, e.g., changes in the EBL upper limits caused by the exclusion of individual data or methods in the EBL upper limit calculations (see ME12 for details).


\section{Discussion and conclusions}\label{Sec:DiscussionConclusions}

As discussed above, recent limits on the EBL density derived from HE/VHE spectra of distant sources constrain its density to a level close to the lower limits derived from integrated galaxy counts (ME12). This implies that the bulk of the EBL density is produced by stellar emission (and stellar emission reprocessed by dust). The EBL density is, thus, directly connected to the star formation history as determined by the SFRD.

When converting these EBL limits into constraints on the SFRD via EBL modeling a peak SFRD lower then the one derived from instantaneous star formation tracers is implied. While the tension between the SFRD from instantaneous tracers and the constraints derived from the EBL limits for the fiducial model adopted in this study is not overly strong ($\sim1\,\sigma$) it must be noted that the parameters of the EBL modeling have considerable uncertainties. For the study presented in this paper, they have been fixed to the most conservative values in the sense that they produce the weakest constraints (i.e. the lowest EBL densities). Consequently, the tension will increase when the parameters are varied from their fiducial values, as it can be expected in the case of, e.g., the metallicity or possibly the IMF.

One possibility to weaken this tension is to adopt an IMF which produces different luminosity outputs for the instantaneous and the late-time SFR tracers, which would imply a mid-heavy ($1-8\,\text{M}_\odot$) IMF \citep[][]{fardal:2007a}. Such an IMF could also explain the difference between the SFRD predicted by instantaneous star formation tracers in comparison to the one derived from the evolution of the stellar mass, as traced by old stellar populations \citep[][]{wilkins:2008a}. The main problem with such a mid-heavy IMF is that it is in contradiction with the finding of \citet{vandokkum:2010a} and (maybe to a lesser degree) \citet{cappellari:2012a}, which both require additional contribution from low mass stars  with masses $<0.5\,\text{M}_\odot$.

While, overall, the constraints on the EBL density are robust (see the discussion in ME12), in the FIR wavelength regime they depend on a single measurement. When removing these data from the sample the EBL limits in the MIR to FIR are significantly weakened (see Figure~7d in ME12). Consequently, this would enable one to weaken the limits on the SFRD derived in this study by adopting a stronger dust attenuation (a higher value of $E(B-V)$, see Figure~\ref{Fig:EBLModel2}).

Other sources, e.g., active galactic nuclei or stars in the early universe, do also contribute to the overall EBL density. For example, \citet{dominguez:2011a} find a contribution of 6 to 13\,\% to the bolometric EBL for AGN-type galaxies. Including such contributions in the modeling and, thereby, increasing the resulting EBL density for a given model, would further strengthen the limits.

In this study, only upper limits on the SFRD have been derived. In principal, with a reasonably well constrained EBL density through upper \emph{and} lower limits, it is possible to derive a best fit SFRD. The main problem for such an ansatz is the insufficient knowledge about many of the parameters in the EBL modeling, e.g., the metallicity evolution, the dust attenuation etc. There is promising progress in reducing the uncertainties, e.g., from the observations of gamma-ray burst afterglows which provide a direct view into the environment of stellar formation at higher redshifts \citep[e.g.][]{zafar:2011a, schady:2012a}, but these measurements are often subject to strong observational biases and the extrapolation needed for a global modeling of the SF are still large.

Upgrades of the current-generation VHE instruments (\HeII, \MaII) are being commissioned specifically targeting the sub-VHE ($\text{E} < 100\,\text{GeV}$) regime, which is key for future VHE EBL studies. Currently under development, with deployment starting as early as 2014, is the Cherenkov Telescope Array \citep[CTA, see ][]{cta:2010:conceptionaldesignreport}. CTA will deliver a ten-times improved sensitivity and extended energy coverage over current generation VHE instruments. EBL studies are one of the major science drivers for CTA and a precise measurement of the EBL density and its evolution from VHE observations will be in reach \citep{raue:2010a}. In the ultra-high gamma-ray energy regime ($E>10\,\text{TeV}$) the non-imaging ground-based detector system SCORE is under development enabling a new view on the MIR to FIR EBL \citep{tluczykont:2011a}.


\appendix

\section{IMF conversion factor}\label{Sec:IMFSalpToChab}

Assumptions about the IMF are necessary to convert measured quantities like luminosities into specific star formation rates and, thereby, the SFRD (see, e.g., HB06). For most quantities, like the IR or the radio luminosity, conversion factors have been calculated assuming a canonical Salp IMF \citep{kennicutt:1998a}. As discussed in Section~\ref{Sec:SSPIMF}, the Chab IMF provides a reasonable description of the average IMF in the local universe. In the following, a scaling factor $f_\text{IMF}$ is derived to convert SFRDs calculated for a Salp IMF $\rho_\ast^\text{Salp}$ to a Chab IMF $\rho_\ast^\text{Chab}$:
\begin{equation}
\rho_\ast^\text{Chab} = \rho_\ast^\text{Salp} / f_\text{IMF} \; .
\end{equation}
For this, the specific luminosities ratios for different SSP spectra from BC03 for a Chab IMF and a Salp IMF are investigated.

\begin{figure}
\centering
\includegraphics[width=0.49\textwidth]{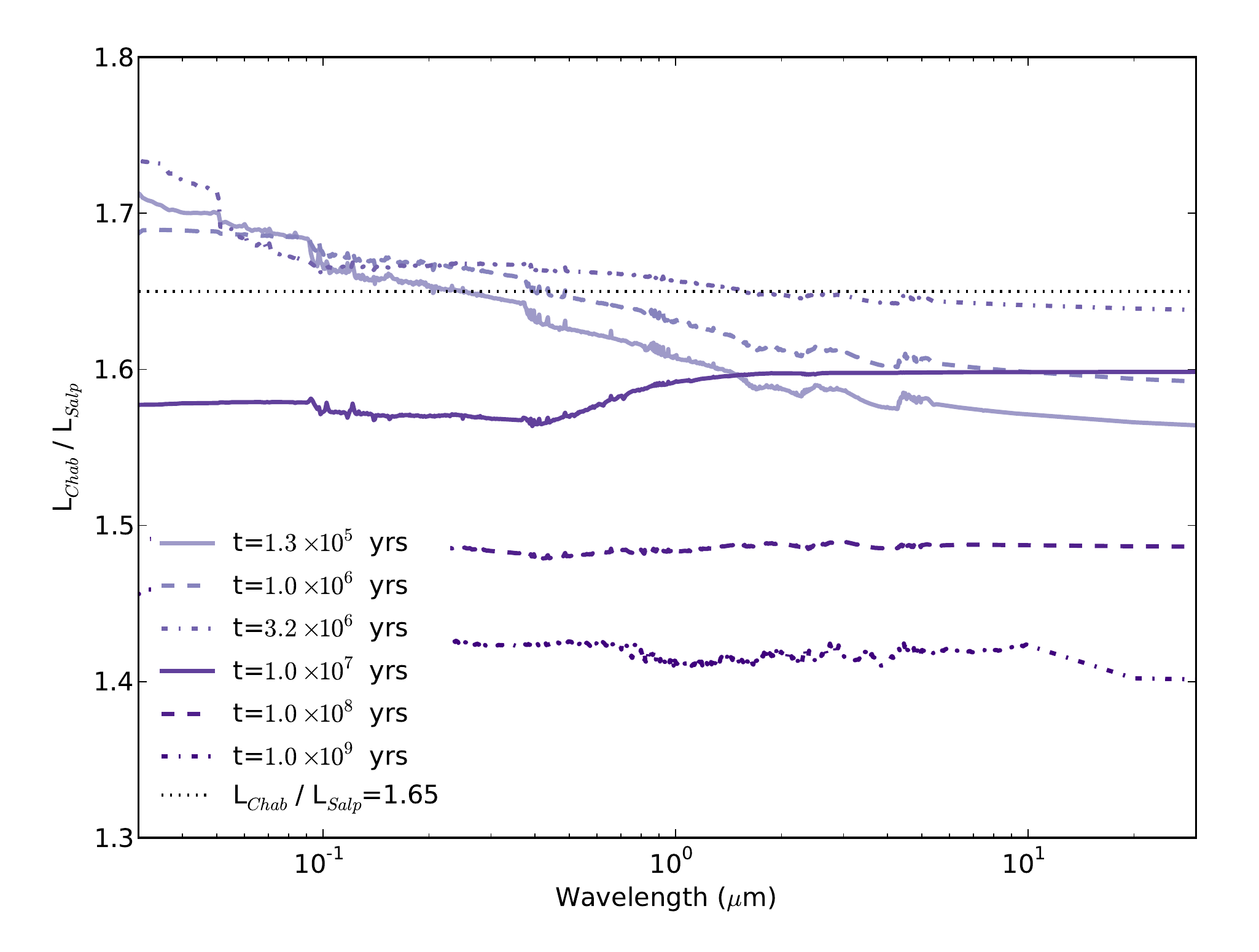}\\
\includegraphics[width=0.49\textwidth]{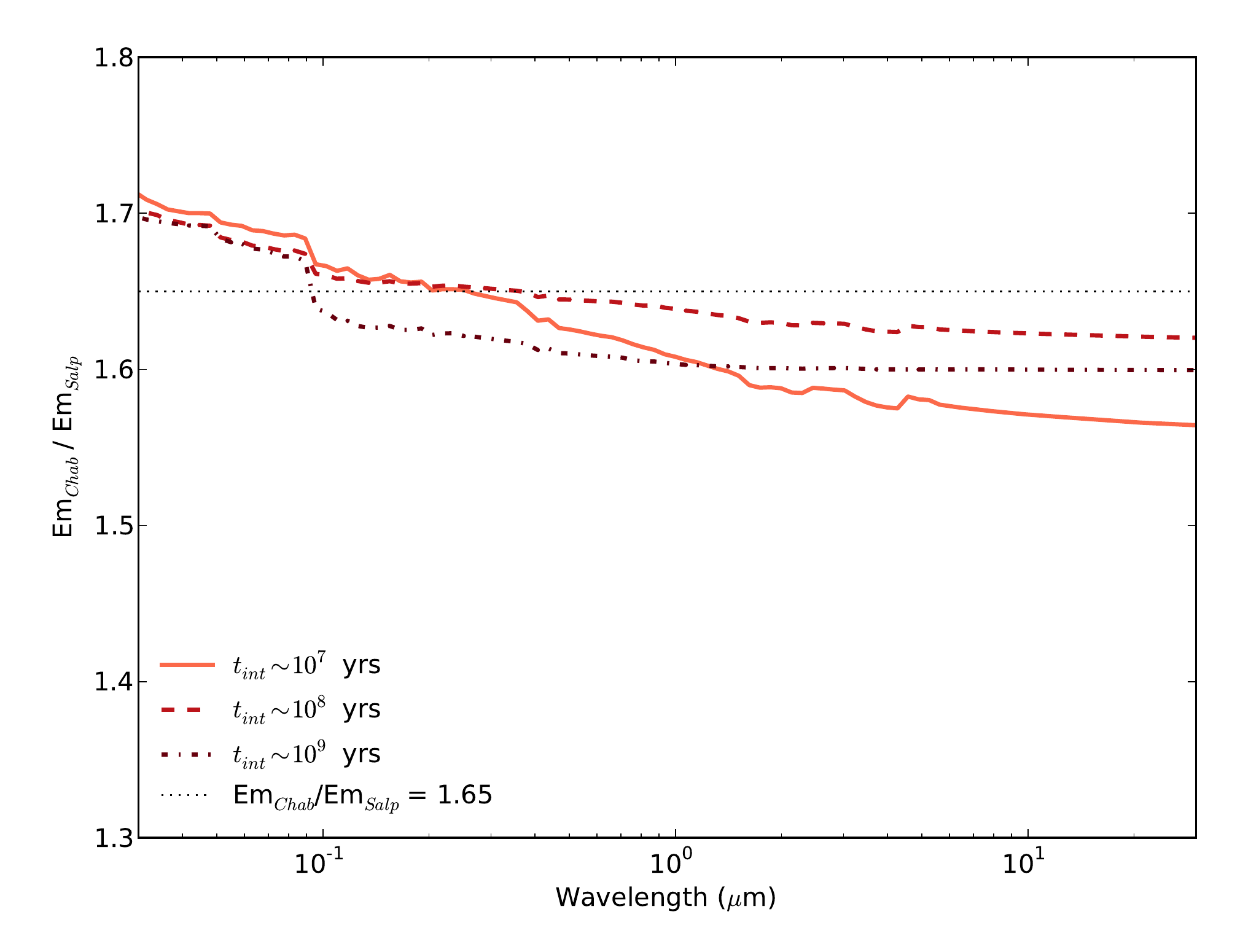}
\caption{\textit{Upper Panel:} SSP luminosity ratio for a Chab and a Salp IMF for different ages of the stellar population. \textit{Lower Panel:} SSP emissivity ratio for a Chab and a Salp IMF for different times assuming a constant SFR.}
\label{Fig:IMFSalpToChab1}
\end{figure}
 
In the top panel of Figure~\ref{Fig:IMFSalpToChab1} the specific luminosities ratios from SSP spectra for a Chab and Salp IMF for different ages are shown. For young ages $<10^7$\,yrs the ratio shows a slightly decreasing slope from $\sim$1.7 at $\sim0.05\,\mu{}m$ to $\sim$1.6 at $\sim20\,\mu{}m$. For later ages the ratio decreases and shows a flat behavior in wavelength. In the lower panel the emissivity for different integration times are shown assuming a constant star formation rate. Here, the ratio shows a similar behavior as for the young age SSP spectra ratio, though with slightly lower values. In HB06 the emissivity value at $0.2\,\mu{}m$ is adopted as a good proxy for the overall scaling. Here this implies an IMF conversion factor of $f_\text{IMF} = 1.65$ (dotted line in the Figures). It can be seen from Figure~\ref{Fig:IMFSalpToChab1} that an IMF conversion factor of $f_\text{IMF} = 1.65$ is in good agreement with the young age SSP spectra and the constant star forming SSPs, though a wavelength dependent spread of about $\pm5\,\%$ remains.
 
\begin{figure}
\centering
\includegraphics[width=0.49\textwidth]{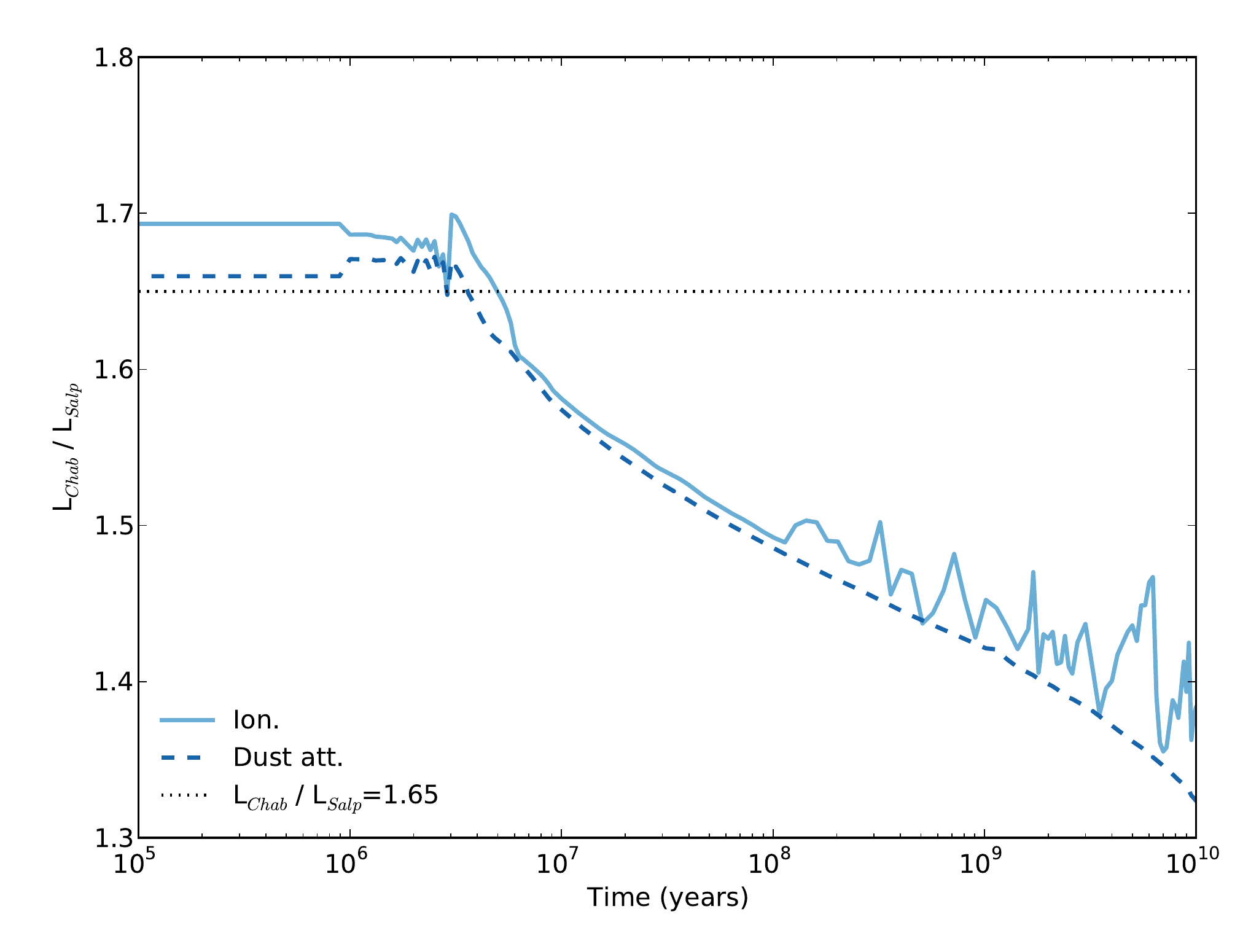}
\caption{Ratio of the integrated ionizing luminosity and luminosity absorbed by dust for SSP models with Chab and  Salp IMF.}
\label{Fig:IMFSalpToChab2}
\end{figure}
 
{
\citet{horiuchi:2009a} derive a slightly higher conversion factor of $f_\text{IMF} \simeq1.8$ for a Kroupa to a Salp IMF using the \textsc{PEGASE.2} code \citep{fioc:1997a}.
}

For completeness, the ratio of the integrated ionizing luminosity and the luminosity absorbed by dust for the Chab and Salp IMF is shown in Figure~\ref{Fig:IMFSalpToChab2}. These quantities are directly related to the total infrared luminosity (i.e. in the EBL model the IR luminosity is the sum of the dust attenuated plus 50\,\% of the ionizing luminosity). Again, for ages up to $10^7$\,yrs, the ratio lies between 1.6 and 1.7, i.e., $f_\text{IMF} = 1.65$ is a good choice.


\section{Systematic uncertainties in the EBL limits}\label{Sec:SysErrEBLLim}

The estimates for the systematic uncertainties in the EBL limits in ME12 are taken from \citet{mazin:2007a}, who used a similar technique as ME12 to derive limits on the EBL density. The systematic uncertainties are estimated to be $\sim32\,\%$ in the O to NIR and $33-55\,\%$ in the MIR to FIR. About $\sim30\,\%$ of this uncertainty results from the choice of the grid spacing, where a smaller grid spacing (i.e. allowing for sharper features in the EBL SED) leads to a relaxed limit. Additional errors arise from the neglected EBL evolution ($\sim10\,\%$), and the uncertainties in the absolute energy scale of ground-based VHE instruments taken to be $\sim15\%$ ($\sim10-45\,\%$ for the O/NIR and the FIR/MIR, respectively). In the study of ME12 the EBL evolution has been included, therefore this contribution can be neglected. In addition, recent studies show that the uncertainty in the absolute energy scale of ground-based VHE instruments is only at the level of $\sim5\,\%$ \citep{meyer:2010a}. The width of the features in the EBL density produced by the EBL modeling in this paper are significantly larger then the ones used to derive the limits in ME12 and, therefore, the systematic error arising from a the choice of the grid spacing can safely be neglected in this study. It is, therefore, concluded that the conservative systematic error on the EBL upper limits relevant for the study presented in this paper is at most $\sim20\,\%$, resulting only from the remaining uncertainty in the absolute energy scale.

\section*{Acknowledgements}

MR and MM acknowledge support of the Hamburg Cluster of Excellence (LEXI).
The authors thank Dieter Horns and Daniel Mazin for helpful comments
and acknowledge interesting discussions with  Shunsaku Horiuchi, David Sobral, and Alberto Dominguez.
The authors would like to thank the anonymous referee for the careful reading of the manuscript.
This research has made use of NASA's Astrophysics Data System
and of the python modules numpy, scipy, and matplotlib \citep{hunter:2007a:matplotlib}.


\def\Journal#1#2#3#4{{#4}, {#1}, {#2}, #3}
\def\NAT{Nature}
\def\AAA{A\&A}
\def\ApJ{ApJ}
\def\AJ{Astronom. Journal}
\def\Aph{Astropart. Phys.}
\def\ApJS{ApJSS}
\def\MNRAS{MNRAS}
\def\NIM{Nucl. Instrum. Methods}
\def\NIMA{Nucl. Instrum. Methods A}
\def\aj{AJ}%
\def\actaa{Acta Astron.}%
\def\araa{ARA\&A}%
\def\apj{ApJ}%
\def\apjl{ApJ}%
\def\apjs{ApJS}%
\def\ao{Appl.~Opt.}%
\def\apss{Ap\&SS}%
\def\aap{A\&A}%
\def\aapr{A\&A~Rev.}%
\def\aaps{A\&AS}%
\def\azh{AZh}%
\def\baas{BAAS}%
\def\bac{Bull. astr. Inst. Czechosl.}%
\def\caa{Chinese Astron. Astrophys.}%
\def\cjaa{Chinese J. Astron. Astrophys.}%
\def\icarus{Icarus}%
\def\jcap{J. Cosmology Astropart. Phys.}%
\def\jrasc{JRASC}%
\def\mnras{MNRAS}%
\def\memras{MmRAS}%
\def\na{New A}%
\def\nar{New A Rev.}%
\def\pasa{PASA}%
\def\pra{Phys.~Rev.~A}%
\def\prb{Phys.~Rev.~B}%
\def\prc{Phys.~Rev.~C}%
\def\prd{Phys.~Rev.~D}%
\def\pre{Phys.~Rev.~E}%
\def\prl{Phys.~Rev.~Lett.}%
\def\pasp{PASP}%
\def\pasj{PASJ}%
\def\qjras{QJRAS}%
\def\rmxaa{Rev. Mexicana Astron. Astrofis.}%
\def\skytel{S\&T}%
\def\solphys{Sol.~Phys.}%
\def\sovast{Soviet~Ast.}%
\def\ssr{Space~Sci.~Rev.}%
\def\zap{ZAp}%
\def\nat{Nature}%
\def\iaucirc{IAU~Circ.}%
\def\aplett{Astrophys.~Lett.}%
\def\apspr{Astrophys.~Space~Phys.~Res.}%
\def\bain{Bull.~Astron.~Inst.~Netherlands}%
\def\fcp{Fund.~Cosmic~Phys.}%
\def\gca{Geochim.~Cosmochim.~Acta}%
\def\grl{Geophys.~Res.~Lett.}%
\def\jcp{J.~Chem.~Phys.}%
\def\jgr{J.~Geophys.~Res.}%
\def\jqsrt{J.~Quant.~Spec.~Radiat.~Transf.}%
\def\memsai{Mem.~Soc.~Astron.~Italiana}%
\def\nphysa{Nucl.~Phys.~A}%
\def\physrep{Phys.~Rep.}%
\def\physscr{Phys.~Scr}%
\def\planss{Planet.~Space~Sci.}%
\def\procspie{Proc.~SPIE}%


\begin{thebibliography}{}

\bibitem[\protect\citeauthoryear{{Aharonian}, {Akhperjanian}, {Bazer-Bachi}
  et~al.,}{{Aharonian} et~al.}{2006}]{aharonian:2006:hess:ebl:nature}
{Aharonian} F.,  {Akhperjanian} A.~G.,  {Bazer-Bachi} A.~R.,    et~al., 2006,
  \nat, 440, 1018

\bibitem[\protect\citeauthoryear{{Albert}, {Aliu}, {Anderhub}, {Antonelli},
  {Antoranz} et~al.,}{{Albert} et~al.}{2008}]{albert:2008:magic:3c279:science}
{Albert} J.,  {Aliu} E.,  {Anderhub} H.,  {Antonelli} L.~A.,  {Antoranz} P.,
  et~al., 2008, Science, 320, 1752

\bibitem[\protect\citeauthoryear{{Binney} \& {Merrifield}}{{Binney} \&
  {Merrifield}}{1998}]{binney:1998a}
{Binney} J.,  {Merrifield} M.,  1998, {Galactic Astronomy}

\bibitem[\protect\citeauthoryear{{Bouwens}, {Illingworth}, {Franx}, {Chary},
  {Meurer}, {Conselice}, {Ford}, {Giavalisco} \& {van Dokkum}}{{Bouwens}
  et~al.}{2009}]{bouwens:2009a}
{Bouwens} R.~J.,  {Illingworth} G.~D.,  {Franx} M.,  {Chary} R.-R.,  {Meurer}
  G.~R.,  {Conselice} C.~J.,  {Ford} H.,  {Giavalisco} M.,    {van Dokkum} P.,
  2009, \apj, 705, 936

\bibitem[\protect\citeauthoryear{{Bruzual} \& {Charlot}}{{Bruzual} \&
  {Charlot}}{2003}]{bruzual:2003a}
{Bruzual} G.,  {Charlot} S.,  2003, \mnras, 344, 1000

\bibitem[\protect\citeauthoryear{{Cappellari}, {McDermid}, {Alatalo}, {Blitz},
  {Bois} et~al.,}{{Cappellari} et~al.}{2012}]{cappellari:2012a}
{Cappellari} M.,  {McDermid} R.~M.,  {Alatalo} K.,  {Blitz} L.,  {Bois} M.,
  et~al., 2012, ArXiv e-prints

\bibitem[\protect\citeauthoryear{{Chary} \& {Elbaz}}{{Chary} \&
  {Elbaz}}{2001}]{chary:2001a}
{Chary} R.,  {Elbaz} D.,  2001, \apj, 556, 562

\bibitem[\protect\citeauthoryear{{CTA Consortium}}{{CTA
  Consortium}}{2010}]{cta:2010:conceptionaldesignreport}
{CTA Consortium} T.,  2010, ArXiv e-prints

\bibitem[\protect\citeauthoryear{{Dole}, {Lagache}, {Puget}, {Caputi},
  {Fern{\'a}ndez-Conde}, {Le Floc'h}, {Papovich}, {P{\'e}rez-Gonz{\'a}lez},
  {Rieke} \& {Blaylock}}{{Dole} et~al.}{2006}]{dole:2006a}
{Dole} H.,  {Lagache} G.,  {Puget} J.-L.,  {Caputi} K.~I.,
  {Fern{\'a}ndez-Conde} N.,  {Le Floc'h} E.,  {Papovich} C.,
  {P{\'e}rez-Gonz{\'a}lez} P.~G.,  {Rieke} G.~H.,    {Blaylock} M.,  2006,
  \aap, 451, 417

\bibitem[\protect\citeauthoryear{{Dom{\'{\i}}nguez}, {Primack}, {Rosario},
  {Prada}, {Gilmore}, {Faber} et~al.,}{{Dom{\'{\i}}nguez}
  et~al.}{2011}]{dominguez:2011a}
{Dom{\'{\i}}nguez} A.,  {Primack} J.~R.,  {Rosario} D.~J.,  {Prada} F.,
  {Gilmore} R.~C.,  {Faber} S.~M.,    et~al., 2011, \mnras, 410, 2556

\bibitem[\protect\citeauthoryear{{Dwek}, {Arendt}, {Hauser}, {Fixsen},
  {Kelsall}, {Leisawitz}, {Pei}, {Wright}, {Mather}, {Moseley}, {Odegard},
  {Shafer}, {Silverberg} \& {Weiland}}{{Dwek} et~al.}{1998}]{dwek:1998a}
{Dwek} E.,  {Arendt} R.~G.,  {Hauser} M.~G.,  {Fixsen} D.,  {Kelsall} T.,
  {Leisawitz} D.,  {Pei} Y.~C.,  {Wright} E.~L.,  {Mather} J.~C.,  {Moseley}
  S.~H.,  {Odegard} N.,  {Shafer} R.,  {Silverberg} R.~F.,    {Weiland} J.~L.,
  1998, \apj, 508, 106

\bibitem[\protect\citeauthoryear{{Dwek} \& {Krennrich}}{{Dwek} \&
  {Krennrich}}{2005}]{dwek:2005a}
{Dwek} E.,  {Krennrich} F.,  2005, \apj, 618, 657

\bibitem[\protect\citeauthoryear{{Edvardsson}, {Andersen}, {Gustafsson},
  {Lambert}, {Nissen} \& {Tomkin}}{{Edvardsson}
  et~al.}{1993}]{edvardsson:1993a}
{Edvardsson} B.,  {Andersen} J.,  {Gustafsson} B.,  {Lambert} D.~L.,  {Nissen}
  P.~E.,    {Tomkin} J.,  1993, \aap, 275, 101

\bibitem[\protect\citeauthoryear{{Fardal}, {Katz}, {Weinberg} \&
  {Dav{\'e}}}{{Fardal} et~al.}{2007}]{fardal:2007a}
{Fardal} M.~A.,  {Katz} N.,  {Weinberg} D.~H.,    {Dav{\'e}} R.,  2007, \mnras,
  379, 985

\bibitem[\protect\citeauthoryear{{Fazio}, {Ashby}, {Barmby}, {Hora}, {Huang},
  {Pahre}, {Wang}, {Willner}, {Arendt}, {Moseley}, {Brodwin}, {Eisenhardt},
  {Stern}, {Tollestrup} \& {Wright}}{{Fazio} et~al.}{2004}]{fazio:2004a}
{Fazio} G.~G.,  {Ashby} M.~L.~N.,  {Barmby} P.,  {Hora} J.~L.,  {Huang} J.-S.,
  {Pahre} M.~A.,  {Wang} Z.,  {Willner} S.~P.,  {Arendt} R.~G.,  {Moseley}
  S.~H.,  {Brodwin} M.,  {Eisenhardt} P.,  {Stern} D.,  {Tollestrup} E.~V.,
  {Wright} E.~L.,  2004, \apjs, 154, 39

\bibitem[\protect\citeauthoryear{{Fernandez} \& {Komatsu}}{{Fernandez} \&
  {Komatsu}}{2006}]{fernandez:2006a}
{Fernandez} E.~R.,  {Komatsu} E.,  2006, \apj, 646, 703

\bibitem[\protect\citeauthoryear{{Fioc} \& {Rocca-Volmerange}}{{Fioc} \&
  {Rocca-Volmerange}}{1997}]{fioc:1997a}
{Fioc} M.,  {Rocca-Volmerange} B.,  1997, \aap, 326, 950

\bibitem[\protect\citeauthoryear{{Gilmore}}{{Gilmore}}{2012}]{gilmore:2012a}
{Gilmore} R.~C.,  2012, \mnras, 420, 800

\bibitem[\protect\citeauthoryear{{Gould} \& {Schr{\'e}der}}{{Gould} \&
  {Schr{\'e}der}}{1967}]{gould:1967a}
{Gould} R.~J.,  {Schr{\'e}der} G.~P.,  1967, Physical Review, 155, 1408

\bibitem[\protect\citeauthoryear{{Hauser}, {Arendt}, {Kelsall}, {Dwek},
  {Odegard} et~al.,}{{Hauser} et~al.}{1998}]{hauser:1998a}
{Hauser} M.~G.,  {Arendt} R.~G.,  {Kelsall} T.,  {Dwek} E.,  {Odegard} N.,
  et~al., 1998, The Astrophysical Journal, 508, 25

\bibitem[\protect\citeauthoryear{{Hauser} \& {Dwek}}{{Hauser} \&
  {Dwek}}{2001}]{hauser:2001a}
{Hauser} M.~G.,  {Dwek} E.,  2001, Annual Review of Astronomy and Astrophysics,
  39, 249

\bibitem[\protect\citeauthoryear{{Hopkins}}{{Hopkins}}{2004}]{hopkins:2004a}
{Hopkins} A.~M.,  2004, \apj, 615, 209

\bibitem[\protect\citeauthoryear{{Hopkins} \& {Beacom}}{{Hopkins} \&
  {Beacom}}{2006}]{hopkins:2006a}
{Hopkins} A.~M.,  {Beacom} J.~F.,  2006, \apj, 651, 142

\bibitem[\protect\citeauthoryear{{Horiuchi}, {Beacom} \& {Dwek}}{{Horiuchi}
  et~al.}{2009}]{horiuchi:2009a}
{Horiuchi} S.,  {Beacom} J.~F.,    {Dwek} E.,  2009, \prd, 79, 083013

\bibitem[\protect\citeauthoryear{Hunter}{Hunter}{2007}]{hunter:2007a:matplotlib}
Hunter J.~D.,  2007, Computing In Science \& Engineering, 9, 90

\bibitem[\protect\citeauthoryear{{Jelley}}{{Jelley}}{1966}]{jelley:1966a}
{Jelley} J.~V.,  1966, Physical Review Letters, 16, 479

\bibitem[\protect\citeauthoryear{{Karim}, {Schinnerer},
  {Mart{\'{\i}}nez-Sansigre}, {Sargent}, {van der Wel}, {Rix}, {Ilbert},
  {Smol{\v c}i{\'c}}, {Carilli}, {Pannella}, {Koekemoer}, {Bell} \&
  {Salvato}}{{Karim} et~al.}{2011}]{karim:2011a}
{Karim} A.,  {Schinnerer} E.,  {Mart{\'{\i}}nez-Sansigre} A.,  {Sargent} M.~T.,
   {van der Wel} A.,  {Rix} H.-W.,  {Ilbert} O.,  {Smol{\v c}i{\'c}} V.,
  {Carilli} C.,  {Pannella} M.,  {Koekemoer} A.~M.,  {Bell} E.~F.,    {Salvato}
  M.,  2011, \apj, 730, 61

\bibitem[\protect\citeauthoryear{{Kennicutt}
  Jr.}{{Kennicutt}}{1998}]{kennicutt:1998a}
{Kennicutt} Jr. R.~C.,  1998, \araa, 36, 189

\bibitem[\protect\citeauthoryear{{Kistler}, {Y{\"u}ksel}, {Beacom}, {Hopkins}
  \& {Wyithe}}{{Kistler} et~al.}{2009}]{kistler:2009a}
{Kistler} M.~D.,  {Y{\"u}ksel} H.,  {Beacom} J.~F.,  {Hopkins} A.~M.,
  {Wyithe} J.~S.~B.,  2009, \apjl, 705, L104

\bibitem[\protect\citeauthoryear{{Kneiske}, {Mannheim} \& {Hartmann}}{{Kneiske}
  et~al.}{2002}]{kneiske:2002a}
{Kneiske} T.~M.,  {Mannheim} K.,    {Hartmann} D.~H.,  2002, \aap, 386, 1

\bibitem[\protect\citeauthoryear{{Kroupa}}{{Kroupa}}{2001}]{kroupa:2001a}
{Kroupa} P.,  2001, \mnras, 322, 231

\bibitem[\protect\citeauthoryear{{Kroupa}, {Weidner}, {Pflamm-Altenburg},
  {Thies}, {Dabringhausen}, {Marks} \& {Maschberger}}{{Kroupa}
  et~al.}{2011}]{kroupa:2011a}
{Kroupa} P.,  {Weidner} C.,  {Pflamm-Altenburg} J.,  {Thies} I.,
  {Dabringhausen} J.,  {Marks} M.,    {Maschberger} T.,  2011, ArXiv e-prints,
  1112.3340

\bibitem[\protect\citeauthoryear{{Leitherer}, {Ortiz Ot{\'a}lvaro}, {Bresolin},
  {Kudritzki}, {Lo Faro}, {Pauldrach}, {Pettini} \& {Rix}}{{Leitherer}
  et~al.}{2010}]{leitherer:2010a}
{Leitherer} C.,  {Ortiz Ot{\'a}lvaro} P.~A.,  {Bresolin} F.,  {Kudritzki}
  R.-P.,  {Lo Faro} B.,  {Pauldrach} A.~W.~A.,  {Pettini} M.,    {Rix} S.~A.,
  2010, \apjs, 189, 309

\bibitem[\protect\citeauthoryear{{Leitherer}, {Schaerer}, {Goldader},
  {Delgado}, {Robert}, {Kune}, {de Mello}, {Devost} \& {Heckman}}{{Leitherer}
  et~al.}{1999}]{leitherer:1999a}
{Leitherer} C.,  {Schaerer} D.,  {Goldader} J.~D.,  {Delgado} R.~M.~G.,
  {Robert} C.,  {Kune} D.~F.,  {de Mello} D.~F.,  {Devost} D.,    {Heckman}
  T.~M.,  1999, \apjs, 123, 3

\bibitem[\protect\citeauthoryear{{Madau} \& {Pozzetti}}{{Madau} \&
  {Pozzetti}}{2000}]{madau:2000a}
{Madau} P.,  {Pozzetti} L.,  2000, \mnras, 312, L9

\bibitem[\protect\citeauthoryear{{Maraston}}{{Maraston}}{2005}]{maraston:2005a}
{Maraston} C.,  2005, \mnras, 362, 799

\bibitem[\protect\citeauthoryear{{Maurer}, {Raue}, {Kneiske}, {Horns},
  {Els{\"a}sser} \& {Hauschildt}}{{Maurer} et~al.}{2012}]{maurer:2012a}
{Maurer} A.,  {Raue} M.,  {Kneiske} T.,  {Horns} D.,  {Els{\"a}sser} D.,
  {Hauschildt} P.~H.,  2012, \apj, 745, 166

\bibitem[\protect\citeauthoryear{{Mazin} \& {Raue}}{{Mazin} \&
  {Raue}}{2007}]{mazin:2007a}
{Mazin} D.,  {Raue} M.,  2007, \aap, 471, 439

\bibitem[\protect\citeauthoryear{{Meyer}, {Horns} \& {Zechlin}}{{Meyer}
  et~al.}{2010}]{meyer:2010a}
{Meyer} M.,  {Horns} D.,    {Zechlin} H.,  2010, \aap, 523, A2+

\bibitem[\protect\citeauthoryear{{Meyer}, {Raue}, {Mazin} \& {Horns}}{{Meyer}
  et~al.}{2012}]{meyer:2012a}
{Meyer} M.,  {Raue} M.,  {Mazin} D.,    {Horns} D.,  2012, accepted for
  publications in \aap, 542, A59

\bibitem[\protect\citeauthoryear{Nikishov}{Nikishov}{1962}]{nikishov:1962a}
Nikishov A.~I.,  1962, Sov. Phys. JETP, 14, 393

\bibitem[\protect\citeauthoryear{{Panter}, {Jimenez}, {Heavens} \&
  {Charlot}}{{Panter} et~al.}{2008}]{panter:2008a}
{Panter} B.,  {Jimenez} R.,  {Heavens} A.~F.,    {Charlot} S.,  2008, \mnras,
  391, 1117

\bibitem[\protect\citeauthoryear{{Pei}}{{Pei}}{1992}]{pei:1992a}
{Pei} Y.~C.,  1992, \apj, 395, 130

\bibitem[\protect\citeauthoryear{{Raue}, {Kneiske} \& {Mazin}}{{Raue}
  et~al.}{2009}]{raue:2009a}
{Raue} M.,  {Kneiske} T.,    {Mazin} D.,  2009, \aap, 498, 25

\bibitem[\protect\citeauthoryear{{Raue} \& {Mazin}}{{Raue} \&
  {Mazin}}{2010}]{raue:2010a}
{Raue} M.,  {Mazin} D.,  2010, Astroparticle Physics, 34, 245

\bibitem[\protect\citeauthoryear{{Raue} \& {Mazin}}{{Raue} \&
  {Mazin}}{2011}]{raue:2011a}
{Raue} M.,  {Mazin} D.,  2011, in Il Nuovo Cimento C 34 03 {EBL studies with
  ground-based VHE gamma-ray detectors: Current status and potential of
  next-generation instruments}

\bibitem[\protect\citeauthoryear{Robertson, Ellis, Dunlop, McLure \&
  Stark}{Robertson et~al.}{2010}]{robertson:2010a}
Robertson B.~E.,  Ellis R.~S.,  Dunlop J.~S.,  McLure R.~J.,    Stark D.~P.,
  2010, Nature, 468, 49

\bibitem[\protect\citeauthoryear{{Rodighiero}, {Vaccari}, {Franceschini},
  {Tresse}, {Le Fevre}, {Le Brun} et~al.,}{{Rodighiero}
  et~al.}{2010}]{rodighiero:2010a}
{Rodighiero} G.,  {Vaccari} M.,  {Franceschini} A.,  {Tresse} L.,  {Le Fevre}
  O.,  {Le Brun} V.,    et~al., 2010, \aap, 515, A8

\bibitem[\protect\citeauthoryear{{Salpeter}}{{Salpeter}}{1955}]{salpeter:1955a}
{Salpeter} E.~E.,  1955, \apj, 121, 161

\bibitem[\protect\citeauthoryear{Santos, Bromm \& Kamionkowski}{Santos
  et~al.}{2002}]{santos:2002a}
Santos M.,  Bromm V.,    Kamionkowski M.,  2002, \mnras, 336, 1082

\bibitem[\protect\citeauthoryear{{Schady}, {Dwelly}, {Page}, {Kr{\"u}hler},
  {Greiner}, {Oates}, {de Pasquale}, {Nardini}, {Roming}, {Rossi} \&
  {Still}}{{Schady} et~al.}{2012}]{schady:2012a}
{Schady} P.,  {Dwelly} T.,  {Page} M.~J.,  {Kr{\"u}hler} T.,  {Greiner} J.,
  {Oates} S.~R.,  {de Pasquale} M.,  {Nardini} M.,  {Roming} P.~W.~A.,  {Rossi}
  A.,    {Still} M.,  2012, \aap, 537, A15

\bibitem[\protect\citeauthoryear{{Sobral}, {Smail}, {Best}, {Geach}, {Matsuda},
  {Stott}, {Cirasuolo} \& {Kurk}}{{Sobral} et~al.}{2012}]{sobral:2012a}
{Sobral} D.,  {Smail} I.,  {Best} P.~N.,  {Geach} J.~E.,  {Matsuda} Y.,
  {Stott} J.~P.,  {Cirasuolo} M.,    {Kurk} J.,  2012, ArXiv e-prints

\bibitem[\protect\citeauthoryear{{Stecker}, {de Jager} \& {Salamon}}{{Stecker}
  et~al.}{1992}]{stecker:1992a}
{Stecker} F.~W.,  {de Jager} O.~C.,    {Salamon} M.~H.,  1992, \apjl, 390, L49

\bibitem[\protect\citeauthoryear{Tluczykont, Hampf, Horns, Kneiske, Eichler,
  Nachtigall \& Rowell}{Tluczykont et~al.}{2011}]{tluczykont:2011a}
Tluczykont M.,  Hampf D.,  Horns D.,  Kneiske T.,  Eichler R.,  Nachtigall R.,
    Rowell G.,  2011, Advances in Space Research, 48, 1935

\bibitem[\protect\citeauthoryear{{Treu}, {Auger}, {Koopmans}, {Gavazzi},
  {Marshall} \& {Bolton}}{{Treu} et~al.}{2010}]{treu:2010a}
{Treu} T.,  {Auger} M.~W.,  {Koopmans} L.~V.~E.,  {Gavazzi} R.,  {Marshall}
  P.~J.,    {Bolton} A.~S.,  2010, \apj, 709, 1195

\bibitem[\protect\citeauthoryear{{van Dokkum} \& {Conroy}}{{van Dokkum} \&
  {Conroy}}{2010}]{vandokkum:2010a}
{van Dokkum} P.~G.,  {Conroy} C.,  2010, \nat, 468, 940

\bibitem[\protect\citeauthoryear{{V{\'a}zquez} \& {Leitherer}}{{V{\'a}zquez} \&
  {Leitherer}}{2005}]{vazques:2005a}
{V{\'a}zquez} G.~A.,  {Leitherer} C.,  2005, \apj, 621, 695

\bibitem[\protect\citeauthoryear{{Wilkins}, {Trentham} \& {Hopkins}}{{Wilkins}
  et~al.}{2008}]{wilkins:2008a}
{Wilkins} S.~M.,  {Trentham} N.,    {Hopkins} A.~M.,  2008, \mnras, 385, 687

\bibitem[\protect\citeauthoryear{{Y{\"u}ksel}, {Kistler}, {Beacom} \&
  {Hopkins}}{{Y{\"u}ksel} et~al.}{2008}]{yuksel:2008a}
{Y{\"u}ksel} H.,  {Kistler} M.~D.,  {Beacom} J.~F.,    {Hopkins} A.~M.,  2008,
  \apjl, 683, L5

\bibitem[\protect\citeauthoryear{{Zafar}, {Watson}, {Fynbo}, {Malesani},
  {Jakobsson} \& {de Ugarte Postigo}}{{Zafar} et~al.}{2011}]{zafar:2011a}
{Zafar} T.,  {Watson} D.,  {Fynbo} J.~P.~U.,  {Malesani} D.,  {Jakobsson} P.,
   {de Ugarte Postigo} A.,  2011, \aap, 532, A143

\end{thebibliography}

\end{document}